\documentclass[12pt]{article}
\newcommand{\1}{\begin{equation}}
\newcommand{\2}{\end{equation}}
\newcommand{\ea}{\begin{eqnarray}}
\newcommand{\ee}{\end{eqnarray}}
\newcommand{\bee}{\begin{eqnarray*}}
\newcommand{\eee}{\end{eqnarray*}}

\newcommand{\pd}[2]{{\frac{{\partial}#2}{{\partial}#1}}}

\newcommand{\de}{{\!\rm d}}
\newcommand{\e}{{\rm e}}

\newcommand{\g}{{\!\,=\,\!}}

\newcommand{\ii}{{\rm i}}

\newcommand{\sa}{\left[ \begin{array} {c} }
\newcommand{\se}{\end{array}\right]}

\newcommand{\"}{{\"a}}

\newcommand{\"}{{\"o}}

\usepackage{texdraw}
\usepackage{graphicx}
\begin{document}
\begin{center}
{\Large\bf  Weyl spin-orbit-coupling-induced interactions in uniform and trapped atomic quantum fluids}
\\*[4mm] {\large Reena Gupta$^1$}, {\large G. S. Singh$^{1,\dag}$},  and {\large J{\"u}rgen Bosse$^{2}$}\\*[1.5mm]
{\small $^1$Physics Department, Indian Institute of Technology Roorkee, Roorkee 247 667, India}\\
{\small $^2$Institute of Theoretical Physics, Freie Universit{\"a}t, Berlin 14195, Germany}\\
*[3mm]
\end{center}

\begin{abstract}
We establish through analytical and numerical studies of thermodynamic quantities for noninteracting atomic gases that the isotropic three-dimensional spin-orbit coupling, the Weyl coupling, induces interaction which counters ``effective" attraction (repulsion) of the exchange symmetry present in zero-coupling Bose (Fermi) gas. The exact analytical expressions for the grand potential and hence for several thermodynamic quantities have been obtained for this purpose in both uniform and trapped cases. It is enunciated that many interesting features of spin-orbit coupled systems revealed theoretically can be understood in terms of coupling-induced modifications in statistical interparticle potential. The temperature-dependence of the chemical potential, specific heat and isothermal compressibility for a uniform Bose gas is found to have signature of the incipient Bose-Einstein condensation in very weak coupling regime although the system does not really go in the Bose-condensed phase. The transition temperature in harmonically trapped case decreases with increase of coupling strength consistent with the weakening of the statistical attractive interaction.  Anomalous behavior of some thermodynamic quantities, partly akin to that in dimensions less than two, appears for uniform fermions as soon as the Fermi level goes down the Dirac point on increasing the coupling strength. It is suggested that the fluctuation-dissipation theorem can be utilized to verify anomalous behaviors from studies of long-wavelength fluctuations in bunching and antibunching effects.
\\*[3mm]
\\*[3mm]
\end{abstract}

\noindent\textbf{I. INTRODUCTION}

It has been during last few years that theoretical proposals came up to have synthetic spin-orbit coupling (SOC) in neutral atoms resulting from   laser-induced coupling between internal psuedospin-$\frac{1}{2}$ degree of freedom and the atomic center-of-mass motion, see \cite{dgj:rmp11, xuy:pra12} and references therein. The required process engineers a spatially homogeneous but non-Abelian vector potential and yields a term in the Hamiltonian which has the form of Rashba  or linear Dresselhaus SO coupling. Both these two-dimensional (2D) couplings, appearing in a natural way in solids under separate physical conditions, have been known for a long time \cite{soc-rd} but are being difficult to implement in atomic systems. However, equal Rashba-Dresselhaus, which is 1D and is now also called \cite{zgz:pra13a} the NIST SOC due to its first implementation by Spielman's group at the NIST \cite{ljs:nat11},  has been realized experimentally in bosonic \cite{ljs:nat11, soc-bose} as well as fermionic \cite{soc-fermi, fhm:pra13} atomic gases. Recently, methods to generate 3D analogues of the Rashba type have also been suggested \cite{ajg:prl12, lzw:prb12}. The proposed 3D couplings for cold atoms could be anisotropic or rotationally invariant. The latter case has been christened by Anderson \emph{et al.} \cite{ajg:prl12} as the Weyl SOC which is, in fact, an \emph{aide m{\'e}moire} to the Weyl fermions \cite{weyl-fermion} that obey Dirac-like equation \cite{mag:05}.

Theoretical studies have been accomplished for a variety of physical phenomena, see references in reviews \cite{dgj:rmp11, zlc:jpb13, reviews-soc}, considering variations in Zeeman field, trapping potential, optical-lattice parameters, detuning and s-wave scattering length $a_{\rm{s}}$ apart from the SO coupling strength. The significant predictions for bosonic systems are that the SOC makes them behave in ways that have no known analogues in conventional condensed-matter systems, namely the half-vortex phase \cite{half-vortex} and the striped superfluid phase \cite{striped}. Also, unconventional BEC beyond ``no-node" theorem \cite{Fey:72} has been analyzed in \cite{zlc:jpb13, no-node} and \emph{Zitterbewegung} oscillations have been observed experimentally \cite{qu5:pra13}. Some of the salient features which have emerged in studies of fermions are nontrivial topological order \cite{gtz:prl11, nontrivial}, enhanced pairing field \cite{ajg:prl12, gtz:prl11, vys:prb11, hjl:prl11, yuz:prl11} and exotic pairing states \cite{zgz:pra13a}, possible existence of Majorana fermions \cite{majorana} and Fulde-Ferrell-Larkin-Ovchinnikov  phase \cite{zgz:pra13a, fflo}, and existence of Dirac cones in honeycomb optical lattices \cite{zhu6:pra13}. These aspects have provided additional impetus to the study of ultracold atoms and matter waves which have been at the forefront for the last two decades to mimic many physical phenomena \cite{reviews-cold}.

It may be recalled in the context of many-body aspects that quantum gases have statistical interactions meaning thereby that ideal bosons (fermions) have ``effective" interparticle attraction (repulsion) as compared to a Boltzmann gas as a consequence of their total wave function being symmetric (antisymmetric) under exchange of a pair of particles, see, for example, Ref. \cite{lal:93}. Thus the two kinds of quantum particles which have either integral or half-odd-integral spin had earlier been considered to be distinct \cite{soc:note4} both at the single-particle level and the many-particle level. However, it is noteworthy in the context of artificial SO coupling that pseudospin-$\frac{1}{2}$ bosons have not only been conceptualized but also experimentally realized. Although the concept is not in conflict with the Pauli spin-statistics theorem \cite{pau:pr40} because the symmetry operations in the pseudospin space are not related to real-space rotations but it implies that bosons and fermions could be made identical at the single-particle level. Despite availability of enormous literature involving all kinds of complexities in the field, a basic question that how does the SOC modify the many-body effect mentioned above is not known. The main purpose of this paper is to address this aspect both in uniform and harmonically trapped gases. The closer analysis of interesting features found in some of the theoretical studies enumerated below suggests that \emph{the SO coupling weakens the statistical interactions in bosons as well as fermions} and we establish it through study of the thermodynamic properties.

It is a textbook result for noninteracting bosons that there is Bose-condensed phase at $T<T_\mathrm{C}$ for the 3D gas, at $T=0$ for the 2D gas, and no condensation at all in the 1D gas. On the other hand, it has recently been pointed out for the Rashba-coupled noninteracting gas that there is no condensation at all in the 2D case \cite{cuz:pra13} and $T_\mathrm{C}$ goes down to zero in the 3D case \cite{hul:pra12, ozb:prl12} consequent upon the single-particle density of states (DOS) acquiring in the low-energy region behavior of reduced spatial dimensionality. The complete destruction of 3D condensate by the Weyl coupling at all $T$, including at $T=0$, in the noninteracting gas has been pointed out in Refs. \cite{cuz:pra13, hey:pra11} to be due to appearance of coupling-weighted 1D DOS. The enhanced single-particle DOS at the Fermi surface due to SOC has been utilized to understand also that the two-body bound state appears in an SO-coupled Fermi gas on the BCS side ($a_s<0$) of the Feshbach resonance howsoever weak the attraction might be \cite{ajg:prl12, vys:prb11} and that the interaction effect gets magnified resulting into increased fermionic-superfluid transition temperature \cite{yuz:prl11}. Also, there is BCS-BEC crossover on increasing SOC while keeping $a_s$ fixed even for $a_s \to  0^-$ \cite {vzs:prb11}.  Cui and Zhou \cite{cuz:pra13}, and Zhou and cui \cite{zhc:prl13} have studied the effect of SOC on an interacting bosonic system and have found that both Rashba and 3D couplings create condensate depletion. Zheng et al. \cite{zyc:jpb13} have discussed shift in the BEC transition temperature in an SO-coupled interacting Bose gas.  Han and S$\acute{a}$ de Melo \cite{has:pra12} have arrived at the  conclusion that the Rashba  makes the Pauli pressure to decrease with increase in the coupling strength even if $a_s$ be kept fixed.  Gong \emph{et al}. \cite{gtz:prl11} have shown that the fermionic superfluidity destroyed by the Zeeman field above a critical value can be restored by a finite SOC without changing $a_s$. Yu \cite{yu:pra12} has noticed that the SOC enhances the short-distance correlation length in Fermi gases.

Exact analytical expressions for grand potentials of Weyl SO-coupled noninteracting systems in uniform as well as trapped systems have been obtained in unified forms introducing a parameter $\eta$ to represent Bose ($\eta = +1$), Fermi ($\eta = -1$) and Boltzmann ($\eta = 0$) gases. The analytical expressions following therefrom for various thermodynamic quantities of the uniform system have universal character for a fixed density, when $\varepsilon_F^0$ serves as an energy scale, as they depend ultimately only on two dimensionless physical parameters: reduced temperature $\tilde T=T/T_F^0$ and reduced SO coupling strength $\tilde\kappa=\kappa/k_F^0$ with $\varepsilon_F^0$, $T_F^0$ and $k_F^0$ being respectively the Fermi energy, Fermi temperature and Fermi wavevector for an ideal gas without SOC. The graphical depiction of the pressure as a function of SOC strength for fixed temperature and density shows that ``effective" interactions become less attractive (repulsive) in uniform as well as trapped bosons (fermions), a result which is corroborated by plots of the isothermal compressibility and analytically in uniform gases by the virial expansion of the equation of state (EOS).

It may be noted that
some of the thermodynamic properties of SO-coupled systems had been studied earlier by He and Yu \cite{hey:pra11}, who found a few interesting results. But their studies were restricted to the uniform gases, and also their main focus was in high- and low- temperature regimes wherein analytical expressions could be obtained. Our studies cover entire range of physical parameters not only for uniform gases but for the trapped systems as well and we retrieve their results as limiting cases of our expressions. Also, we provide a sound theoretical basis for (i) the observation by Han and S$\acute{a}$ de Melo \cite{has:pra12}, while studying theoretically BCS-BEC crossover in a balanced mixture of 3D ultracold fermions, that the Pauli pressure decreases and consequently the isothermal compressibility increases on increase in the Rashba coupling strength, and (ii) the finding of Ghosh \emph{et al}. \cite{ghosh3:pra11} regarding decrease in Pauli pressure with increase of 3D SOC in uniform systems while studying trapped fermions.  Furthermore, we find signature of incipient Bose-condensation for very weak SO coupling in a uniform gas although there is really no condensation due to presence of coupling-weighted 1D  DOS. We observe anomalous behavior, typical of spatial-dimensional crossover, in some thermodynamic quantities of uniform fermions when the Fermi level crosses the Dirac point on varying the coupling strength. Of course, the physics of harmonically trapped gases is more interesting as the phenomenon of BEC destroyed by the SOC in 3D Bose gas is restored by trapping even in the noninteracting case; our expressions offer opportunity for the study of complete thermodynamics of such systems.

The paper is structured as follows. Section II starts with an overview of basic aspects pertaining to the Weyl SO-coupled Hamiltonian for noninteracting atomic quantum gases, and the behavior of the resulting DOS is described for both uniform and trapped gases. The analytical expressions for grand potentials involving polylogarithmic functions are presented for these systems in Sections IIIA and IIIB, respectively, and exact analytical expressions for various thermodynamic quantities are developed in both the cases.  Computed results are depicted graphically in Section IV wherein detailed discussions are also given. Section V gives summary and outlook of our work. Appendix A discusses some aspects pertaining to plots of DOS in uniform gases, Appendix B contains derivation for the expression of the isochoric specific heat, and detailed discussion of the fermionic chemical potential is given in Appendix C.\\

\noindent\textbf{II. WEYL SPIN-ORBIT-COUPLED SYSTEMS}\\
\noindent\textbf{A. Basic Aspects}

We consider a system of cold neutral atoms, bosons or fermions, having multilevel hyperfine states and moving in the presence of spatially varying laser fields. The coupling induced via two-photon Raman transitions linking four internal levels in a tetrahedral topology, as described in \cite{ajg:prl12}, gives rise to an effective non-Abelian vector potential $\hat{\textbf{A}}=-\hbar\kappa\hat\mathbf{\sigma}$, where $\kappa$ is the isotropic coupling strength and $\mathbf{\hat\mathbf{\sigma}}=\left(\sigma_x, \sigma_y, \sigma_z\right)$ is the conventional Pauli spin operator in the pseudospin space. Hence the single-particle effective Hamiltonian of $(\uparrow, \downarrow)$ pseudospin-$\frac{1}{2}$  particles would become $\hat{\cal H}=\left(1/2m\right)\left(\hat{\textbf{p}}I-\hat{\textbf{A}}\right)^2$ which one may take, without any loss of generality, in the form
\1
\label{hamiltonian}
\hat{\cal H}=\frac{1}{2m} \left(\hat{\textbf{p}}^2 +\hbar^2\kappa^2\right) I +
v\mathbf{\hat{\mathbf{\sigma}}}\cdot\hat{\textbf{p}}.
\2
Here $v=\hbar\kappa/m$ is the coupling strength having unit of velocity, $I$ is the $2\times2$ identity matrix, and $\hat{\textbf{p}}$ is the momentum operator.

It may be noted that the atoms have acquired psuedospin-$\frac{1}{2}$ degree of freedom, irrespective of the system without coupling being fermionic or bosonic, which has got coupled with the atomic center-of-mass motion. If the inconsequential constant term be not retained in Eq. (\ref{hamiltonian}), the resulting isotropic coupling term $\hat{\cal H}_{\rm{SO}}=v\hat{\mathbf{\sigma}}\cdot\hat{\textbf{p}}$ is a natural symmetric extension of the Rashba coupling to 3D and has the form of the Weyl Hamiltonian \cite{mag:05}. One has thus the Weyl SO coupling, the terminology coined in \cite{ajg:prl12}. Furthermore, the Hamiltonian $\hat{\cal H}$ possesses complete rotational symmetry, the highest such a symmetry among all types of SO couplings, as a consequence of the commutation relation $\left[\hat{\textbf{J}},  \hat{\cal H}\right]=0$, where $\hat{\textbf{J}}= \hat{\textbf{L}}+\frac{1}{2} \hbar\hat{\mathbf{\mathbf{\sigma}}}$ is the total angular momentum operator with $\hat{\textbf{L}}$ as atomic center-of-mass orbital angular momentum operator. Also, the Hamiltonian respects the time-reversal symmetry even when $\kappa\ne 0$ but the Galilean invariance and the parity conservation are broken. Various aspects of the simplicity inherently present in the Weyl-coupled Hamiltonian for isotropically trapped gases described by Anderson and Clark \cite{anc:jpb13} apply here in case of uniform gases too.

The eigenvalues $\pm 1$ of the helicity operator $\mathbf{\hat{\mathbf{\sigma}}}\cdot\hat{\textbf{p}}/|\textbf{p}|$ are good quantum numbers and the Hamiltonian in Eq. (\ref{hamiltonian}) can be diagonalized yielding eigenenergies
\1
\label{dispersion}
\varepsilon_{\pm}\left(\textbf{k}\right)=\frac{\hbar^2}{2m}\left(\rm{k}\pm\kappa\right)^2
\2
which correspond to two orthogonal eigenstates $|\textbf{k}^{+}\rangle={u}_{\rm{\bf{k}}}|\textbf{k}_{\uparrow}\rangle +{v}_{\rm{\bf{k}}}\e^{\ii\phi_{\rm{\bf{k}}}}|\textbf{k}_{\downarrow}\rangle$ and $|\textbf{k}^{-}\rangle={u}_{\rm{\bf{k}}}|\textbf{k}_{\downarrow}\rangle -{v}_{\rm{\bf{k}}}\e^{-\ii\phi_{\rm{\bf{k}}}}|\textbf{k}_{\uparrow}\rangle$ with  ${u}_{\rm{\bf{k}}}=\left[\left(1+\rm{k_z}/|\rm{\bf{k}}|\right)/2\right]^{1/2}$, ${v}_{\rm{\bf{k}}}=\left[\left(1-\rm{k_z}/|\rm{\bf{k}}|\right)/2\right]^{1/2}$ and $\phi_{\rm{\bf{k}}}=\arg\left(\rm{k_x}+\ii\rm{k_y}\right)$.
The system has spherical manifold of zero-energy ground states and there is a 3D Dirac point at $\textbf{k}=0$ which is robust since the presence of all the three spin matrices in $\hat{\cal H}_{\rm{SO}}$ ensures that a homogeneous Zeeman field cannot lift up the degeneracy at this point. The point is located in energy space at $\varepsilon_+=\varepsilon_-=\hbar^2\kappa^2/2m$ which is the spin-orbit coupling energy $\varepsilon_{\mathrm{so}}$. Its significance gets highlighted below when we discuss SO-induced dimensional reduction, and again in Section IV when we discuss appearance of anomalous behavior in some thermodynamic quantities of fermions as soon as the Fermi level goes down this point on increasing coupling strength.

Equation (\ref{dispersion}) implies that the system possesses two constant energy surfaces (CESs) in $\textbf{k}$-space, both being spherical having radii $\rm{k}_\pm= k_0\mp\kappa$, where $k_0=\left(2m\varepsilon_0 /\hbar^2\right)^{1/2}$ with $\varepsilon_0=\varepsilon_+=\varepsilon_-$ having a chosen constant value. For $\kappa=0$, two spheres have the same radii $(\rm{k}_+= \rm{k}_-=k_0)$ but as soon as the SO coupling is switched on, we have two concentric spheres with $k_+<k_0$ and $k_->k_0$, i.e. the volume of the CES corresponding to the positive helicity branch gets decreased whereas that of the negative branch gets increased. When $\kappa$ would increase, the smaller(larger) volume CES would shrink(expand) and for $\kappa= k_0$, the + branch CES would reduce to a point. This sequence of interesting behavior together with that for $\kappa>k_0$ unveils for fermions the so-called Fermi-surface-topology transition described in \cite{vzs:prb11, svg:cusc12}.

The Hamiltonian (\ref{hamiltonian}) can be generalized for a trapped gas to have the semiclassical form \cite{hul:pra12, anc:jpb13} wherein the trapping potential $U(r)$ would simply get added on the right hand side in Eq. (\ref{hamiltonian}) and the counterpart of Eq.(\ref{dispersion}) would become the semiclassical energy spectrum given by
\ea
\label{dispersion-semi}
\varepsilon_{\pm}\left(k, r\right)=\frac{\hbar^2}{2m}\left(\rm{k}\pm\kappa\right)^2 + U(r)
\ee
in the $6$-dimensional phase space ($\mathbf{k}, \mathbf{r}$). The motion of the particle would now get restricted to the region of space wherein $U(r)$ is less than the particle's energy $\varepsilon$. This introduces the concept of the effective volume as described in \cite{bagnato3:pra87} and can be expressed for the 3D systems as
\ea
V_{3D}^*(\varepsilon)=\int \Theta\left(\varepsilon-U(r)\right)~\de{\mathbf{r}}=\frac{4\pi}{3}r_{\mathrm{max}}^3
\ee
where $\Theta(x)$ is the Heaviside step function and $r_{\mathrm{max}}$ is the solution of the equation $U(r_{\mathrm{max}})=\varepsilon$.

\noindent\textbf{B. Density of States}

If the energy-level separation be much smaller than the thermal energy $k_\mathrm{B} T$, which is usually the situation under experimental conditions for cold atoms, the eigenenergies can be taken to be quasi-continuous and the Thomas-Fermi semiclassical approximation would become applicable. Consequently, the density of states (DOS) for each branch can be calculated using the expression
\ea
\label{dos-partial}
G_{\pm}(\varepsilon)&=&\frac{1}{(2\pi)^3}\int\int \de ^3 r~\de ^3 k~ \delta\left(\varepsilon-\varepsilon_{\pm}\left(k,r\right)\right).
\ee
Here $\delta(x)$ is the Dirac delta function and $\de ^3r~\de ^3k/(2\pi)^3$ is the number of quantum states in the phase-space elementary volume $\de ^3r~\de ^3k$ with $(2\pi)^3$ being the volume occupied by a single quantum state.

In order to evaluate the integrals appearing in Eq.(\ref{dos-partial}), we require the explicit form for $U(r)$. We choose this isotropic trapping potential to have the generic power-law behavior: $U(r)=U_0 (r/\ell_0)^{\gamma}$ with $U_0$ and $\ell_0$ as constants. The value of the exponent $\gamma$ defines the nature of the potential; $\gamma=2$ gives the important case of the harmonic trapping and $\lim_{\gamma\to\infty} U(r)$ gives the spherical potential well of radius $\ell_0$: $U(r)=0$ for $r<\ell_0$ and $U(r)=\infty$ for $r>\ell_0$. Thus the uniform gas corresponds to the limiting situation when $\gamma\to\infty$ and $\ell_0\to\infty$.

The evaluation requiring the use of the integral \cite{gradeshteyn2:book07}
\ea
\int_{0}^ {\pi/2}\sin^{\mu-1}\theta\cos^{\nu-1}\theta~\de\theta
&=&\frac{1}{2}\frac{\Gamma\left(\frac{\mu}{2}\right)\Gamma\left(\frac{\nu}{2}\right)}{\Gamma\left(\frac{\mu}{2}+\frac{\nu}{2}\right)}; \Re\mu>0, \Re\nu>0
\ee
has ultimately yielded the expression for the total DOS, $G(\varepsilon)\equiv G(\varepsilon,\kappa,\gamma)=G_{+}(\varepsilon)+G_{-}(\varepsilon)$, in the form
\ea
\label{dos-general}
G(\varepsilon)&=&\left(\frac{4\pi\ell_0^3}{3}\right)\left(\frac{\varepsilon}{U_0}\right)^{\frac{3}{\gamma}}\Gamma\left(\frac{3}{\gamma}+1\right)
\left[\frac{\Gamma\left(\frac{3}{2}\right)\rho^{0}_{3D}\left(\varepsilon\right)}{\Gamma\left(\frac{3}{\gamma}+\frac{3}{2}\right)}
+ \frac{\kappa^2}{2\pi}\frac{\Gamma\left(\frac{1}{2}\right)\rho^{0}_{1D}\left(\varepsilon\right)}
{\Gamma\left(\frac{3}{\gamma}+\frac{1}{2}\right)}\right],
\ee
where
\ea
\label{dos-dDim}
\rho^0_{d}(\varepsilon)=\frac{1}{2^{d-1}\pi^{d/2}\Gamma(d/2)}\left(\frac{2m}{\hbar^2}\right)^{d/2}\varepsilon^{d/2-1}~\Theta(\varepsilon)
\ee
is the  DOS per unit volume for both spin orientations of a d-dimensional uniform system of spin-$\frac{1}{2}$ particles without SOC, and $\Gamma(\nu)$ is the gamma function. Thus we have obtained a quite general expression in a neat analytical form for 3D systems and we now examine its validity in known special cases.

The expression, as given in \cite{salasnich1:jmp00}, for the DOS of ideal quantum gases under the influence of a power-law external potential but without SOC is recovered in 3D case on substituting $\kappa=0$ and $\ell_0=1$ in Eq.(\ref{dos-general}). Also, for the most important case of the harmonic trapping potential, $U(r)=\frac{1}{2}m\omega^2r^2$, we get
\ea
\label{dos-harmonic}
G_H(\varepsilon)&=& \frac{\varepsilon^2}{\hbar^3\omega^3}\left(1+4~\frac{\varepsilon_{\rm{so}}}{\varepsilon}\right)\Theta(\varepsilon)
\ee
whose comparison with the well known result for harmonically trapped Bose gas without SOC reveals the presence of an extra factor 2 here because we have considered the pseudospin-$\frac{1}{2}$ bosons.
In order to get the result for SO-coupled uniform gases, however, we first obtain from Eq.(\ref{dos-general}) the DOS per unit effective volume and then take the limit $\gamma\to\infty$. Since $r_{\mathrm{max}}=\ell_0(\varepsilon/U_0)^{1/\gamma}$, the procedure gives
the DOS $\rho\left(\varepsilon\right)\equiv\rho_{3D}\left(\varepsilon,\kappa\right)$ as
\ea
\label{dos}
\rho\left(\varepsilon\right)&=&\rho^{0}_{3D}\left(\varepsilon\right)\left(1+\frac{\varepsilon_{\rm{so}}}{\varepsilon}\right),
\ee
a rewritten form of that given in Ref. \cite{hey:pra11}.
The detailed behavior of this DOS is analyzed in Appendix \ref{appen-dos} wherein it is found that $\rho\left(\varepsilon\right)$ possesses a minimum at $\varepsilon=\varepsilon_{\rm{so}}$ with $\varepsilon_{\rm{so}}=\hbar^2\kappa^2/2m$ being the SO coupling energy which corresponds to the location of the Dirac point in a uniform gas.
The presence of SOC-weighted 1D DOS in Eq. (\ref{dos}) imparts the system characteristics of reduced spatial dimensionality: quasi-1D (more than 1D) for $\varepsilon<\varepsilon_{so}$ and quasi-3D (less than 3D) for $\varepsilon>\varepsilon_{so}$. Thus \emph{the uniform system undergoes dimensional reduction for large $\kappa$} analogous to that mentioned in Ref. \cite{anc:jpb13} for SO-coupled harmonically trapped gases.

We have the inequality $\rho\left(\varepsilon\right)\ge\rho^{0}_{3D}\left(\varepsilon\right)$ wherein the equality holds for $\varepsilon=\infty$ or $\kappa\g0$. This SOC-induced enhancement in DOS has been utilized in the literature, see for example, in \cite{ajg:prl12, cuz:pra13, hey:pra11, vzs:prb11}, to give plausible reasoning to understand some of the interesting features enumerated in the Introduction. However,  we assert that not only the explanations for these features but also for the additional ones mentioned in the Introduction could be constructed in terms of SOC-induced modifications in the statistical interactions amongst particles of the system described in Sec. IV.

\noindent\textbf{III. ANALYTICAL THERMODYNAMICS}

It could be noted that the process of evaluation for the thermodynamic quantities involves energy integration and hence once the integration has been accomplished, the expressions so obtained cannot be used to get results for the uniform case even if the generic power-law potential be used. Hence we consider here separately both uniform and harmonically trapped gases having Weyl SOC to obtain exact analytical expressions for various thermodynamic quantities.

\noindent\textbf{A. Uniform Gases}

The grand thermodynamic potential $\Omega(\mu,T,V)$ of Bose ($\eta=+1$),  Fermi ($\eta=-1$) and Boltzmann ($\eta=0$) gases can be written in a unified form as
\ea
\label{grandpot-def}
\Omega&=&\frac{1}{\eta\beta}\sum_{\textbf{k},\lambda=\pm}\ln\left[1-\eta z\exp\{-\beta\varepsilon_\lambda\left(\textbf{k}\right)\}\right]
\nonumber\\
&=& \frac{V}{\eta\beta}\int_0^\infty \rho(\varepsilon)\ln\left(1-\eta z\e^{-\beta\varepsilon}\right) \de \varepsilon,
\ee
where $z=\e^{\beta\mu}$ is the fugacity, $\mu$ the chemical potential, $\beta=1/(k_B T)$ the inverse temperature, and $k_B$ the Boltzmann constant. Using Eq.(\ref{dos}) in Eq.(\ref{grandpot-def}), substituting $x=\beta\varepsilon$ and introducing the thermal de Broglie wavelength $\Lambda=\left(2\pi\hbar^2\beta/m\right)^{1/2}$, one gets
\ea
\label{grandpot}
\Omega=\frac{4V}{\sqrt\pi \eta\beta\Lambda^3}\int_0 ^\infty\left(\sqrt x
+\frac{\kappa^2\Lambda^2}{4\pi\sqrt x}\right)\ln\left(1-\eta z\e^{-x}\right) \de x
\ee

Now we use the identity
\1
\nonumber \frac{u}{\e^x - u}=\frac{\de}{\de x}\ln\left(1-u\e^{-x}\right)
\2
in the integral representation \cite{pbm:book90}
\1
\label{polylog}
\rm{Li}_\nu(u)=\frac{u}{\Gamma(\nu)}\int_0 ^\infty \frac{x^{\nu-1}}{\e^x - u} \de x
\2
for the polylogarithm of $\nu$th order with $\nu>0$, and then perform the integration by parts. The process yields finally
\1
\rm{Li}_{\nu+2}(u)=-\frac{1}{\Gamma(\nu+1)}\int_0 ^\infty x^\nu \ln\left(1-u\e^{-x}\right) \de x
\2
with $\nu>-1$. Hence Eq. (\ref{grandpot}) is rewritten in a neat analytical form as
\1
\label{omega-final}
\Omega=-\frac{2V}{\eta\beta\Lambda^3}\left[\rm{Li}_{5/2}\left(\eta z\right)+\phi\left(\kappa\Lambda\right)\rm{Li}_{3/2}\left(\eta z\right)\right]
\2
with $\phi\left(\kappa\Lambda\right)= {\kappa^2\Lambda^2}/{2\pi}$, and can be utilized to discuss complete thermodynamics of the system.

The total number of particles and the entropy given by $N=-\left(\partial\Omega/\partial\mu\right)_{T,V}$ and $S=-\left(\partial\Omega/\partial T\right)_{V,\mu}$ yield expressions
\1
\label{number}
N=\frac{2V}{\eta\Lambda^3}\left[\rm{Li}_{3/2}\left(\eta z\right)+\phi\left(\kappa\Lambda\right)\rm{Li}_{1/2}\left(\eta z\right)\right]
\2
and
\1
\label{entropy}
\frac{S}{k_B} =\frac{V}{\eta\Lambda^3}\left[5\rm{Li}_{5/2}\left(\eta z\right)+3\phi\left(\kappa\Lambda\right)\rm{Li}_{3/2}\left(\eta z\right)\right]-
N\beta\mu,
\2
where we have used the recurrence relation
\1
\label{recurrence}
\frac{\partial}{\partial z}\rm{Li}_\nu(z)=\frac{1}{z}\rm{Li}_{\nu-1}(z)
\2
obtained from Eq.(\ref{polylog}). The relation $\Omega=-PV$ gives pressure
\1
\label{pressure}
P=\frac{2}{\eta\beta\Lambda^3}\left[\rm{Li}_{5/2}\left(\eta z\right)+\phi\left(\kappa\Lambda\right)\rm{Li}_{3/2}\left(\eta z\right)\right].
\2

The internal energy and the Helmholtz free energy can be obtained through Legendre transformations $E=\Omega+TS+\mu N$ and $F=\Omega+\mu N$. Thus
\1
\label{energy}
 E =\frac{V}{\eta\beta\Lambda^3}\left[3\rm{Li}_{5/2}\left(\eta z\right)+\phi\left(\kappa\Lambda\right)\rm{Li}_{3/2}\left(\eta z\right) \right]
\2
and
\ea
\label{free energy}
F =\mu N- \frac{2V}{\eta\beta\Lambda^3}\left[\rm{Li}_{5/2}\left(\eta z\right)+\phi\left(\kappa\Lambda\right)\rm{Li}_{3/2}\left(\eta z\right)\right],
\ee
and the EOS is
\1
\label{EOS}
PV=\frac{2}{3}\left[E+\left(\frac{\partial E}{\partial\ln\kappa}\right)_{\mu}\right],
\2
a form obtained also in Ref. \cite{hey:pra11}. On using Eq.(\ref{number}) in the definition $K_T=-(1/V)\left(\partial V/\partial P\right)_{T,N}=(1/n)\left(\partial n/\partial P\right)_{T,N}=(1/n^2)\left(\partial n/\partial \mu\right)_{T,N}$, the isothermal compressibility takes the form
\1
\label{compress}
K_{\mathrm{T}}=\frac{2\beta}{\eta n^2\Lambda^3}\left[\rm{Li}_{1/2}\left(\eta z\right)+\phi\left(\kappa\Lambda\right)\rm{Li}_{-1/2}\left(\eta z\right)\right].
\2

For $|u|<1$, Eq.(\ref{polylog}) has the series expansion $\rm{Li}_{\nu}\left(u\right)=\sum_{\ell=1}^\infty (u^\ell/\ell^\nu)$.
Hence for $|\eta z|<1$, Eqs.(\ref{number}), (\ref{energy}) and (\ref{pressure}) give
\1
\label{number-series}
N=\frac{2V}{\eta\Lambda^3}\sum_{\ell=1}^\infty\left(\eta z\right)^{\ell}
\left(\frac{1}{\ell^{3/2}}+\phi\left(\kappa\Lambda\right)\frac{1}{\ell^{1/2}}\right),
\2
\1
\label{energy-series}
E=\frac{V}{\eta\beta\Lambda^3}\sum_{\ell=1}^\infty\left(\eta z\right)^{\ell} \left(\frac{3}{\ell^{5/2}}+\phi\left(\kappa\Lambda\right)\frac{1}{\ell^{3/2}}\right)
\2
and
\1
\label{pressure-series}
P=\frac{2}{\eta\beta\Lambda^3}\sum_{\ell=1}^\infty \left(\eta z\right)^{\ell} \left(\frac{1}{\ell^{5/2}}+\phi\left(\kappa\Lambda\right)\frac{1}{\ell^{3/2}}\right).
\2
For bosons and fermions, above three equations yield respectively Eqs. (10), (9) and (8) of Ref. \cite{hey:pra11} valid in the high-temperature limit when fugacity $z<1$. Additionally, if one considers the Boltzmann ($\eta=0$) gas, only the leading term , i.e. $\ell=1$, survives in the expression for $N$, $E$ or $P$ on taking the limit $\eta\to0$.

The evaluation for the specific heat $C_V=\left(\partial E/\partial T\right)_{S,V}$ has been presented in Appendix \ref{appen-spheat} and the final expression is
\ea
\label{specific heat}
\frac{C_V}{N k_B}=\frac{3}{4}\frac{\left[5\rm{Li}_{5/2}\left(\eta z\right)+\phi\left(\kappa\Lambda\right)\rm{Li}_{3/2}\left(\eta z\right)\right]}{\left[\rm{Li}_{3/2}\left(\eta z\right)+\phi\left(\kappa\Lambda\right)\rm{Li}_{1/2}\left(\eta z\right)\right]}~~~~~~~~~~~~~~~~~~~~~~~~~~~~~~~~~~~~~~~~~~~~~\nonumber\\
-\frac{1}{4}\frac{\left[3\rm{Li}_{3/2}\left(\eta z\right)+ \phi\left(\kappa\Lambda\right)\rm{Li}_{1/2}\left(\eta z\right)\right]\left[3\rm{Li}_{3/2}\left(\eta z\right)+\phi\left(\kappa\Lambda\right)\rm{Li}_{1/2}\left(\eta z\right)\right]}{\left[\rm{Li}_{3/2}\left(\eta z\right)+\phi\left(\kappa\Lambda\right)\rm{Li}_{1/2}\left(\eta z\right)\right]\left[\rm{Li}_{1/2}\left(\eta z\right)+\phi\left(\kappa\Lambda\right)\rm{Li}_{- 1/2}\left(\eta z\right)\right]}.
\ee
The evaluation for thermal expansion coefficient $\alpha_{\mathrm{T}}=(1/V)\left(\partial V/\partial T\right)_{P,N}=K_{\mathrm{T}}\left(\partial P/\partial T\right)_{V,N}$ involves some aspects as in case of the evaluation of $C_V$, and we obtain finally
\ea
T\alpha_{\mathrm{T}}&=& \frac{5}{2} \frac{\left[\rm{Li}_{5/2}\left(\eta z\right)+\phi\left(\kappa\Lambda\right)
\rm{Li}_{3/2}\left(\eta z\right)\right]\left[\rm{Li}_{1/2}\left(\eta z\right)+\phi\left(\kappa\Lambda\right)\rm{Li}_{-1/2}\left(\eta z\right)
\right]}{\left[\rm{Li}_{3/2}\left(\eta z\right)+\phi\left(\kappa\Lambda\right)\rm{Li}_{1/2}\left(\eta z\right)\right]^2}\nonumber\\
&&-\phi\left(\kappa\Lambda\right)\rm{Li}_{3/2}\left(\eta z\right)\frac{\left[\rm{Li}_{1/2}\left(\eta z\right)+\phi\left(\kappa\Lambda\right)
\rm{Li}_{-1/2}\left(\eta z\right)\right]}{\left[\rm{Li}_{3/2}\left(\eta z\right)+\phi\left(\kappa\Lambda\right)\rm{Li}_{1/2}\left(\eta z\right)\right]^2}\nonumber\\
&&-\frac{1}{2}\frac{\left[3\rm{Li}_{3/2}\left(\eta z\right)+\phi\left(\kappa\Lambda\right)
\rm{Li}_{1/2}\left(\eta z\right)\right]}{\left[\rm{Li}_{3/2}\left(\eta z\right)+\phi\left(\kappa\Lambda\right)\rm{Li}_{1/2}\left(\eta z\right)\right]}
\ee

It can be noted that on taking the limit $\kappa\to0$, all our expressions with the SOC reduce to the corresponding results for ideal quantum gases in terms of polylogs \cite{lee:09} and to those given in textbooks, see for example \cite{pab:11}, in terms of Bose and Fermi functions. Also, the thermodynamic relations $C_{\mathrm{P}}=C_{\mathrm{V}} + VT\alpha^2_{\mathrm{T}}/K_{\mathrm{T}}$ and $K_{\mathrm{S}}=K_{\mathrm{T}}-VT\alpha^2_{\mathrm{T}}/C_{\mathrm{P}}$, where $K_{\mathrm{S}}$ is the adiabatic compressibility, suggest that the study of complete thermodynamics is possible with the help of the expressions derived above.

\begin{figure}
\begin{center}
\includegraphics[width=80mm,angle=0]{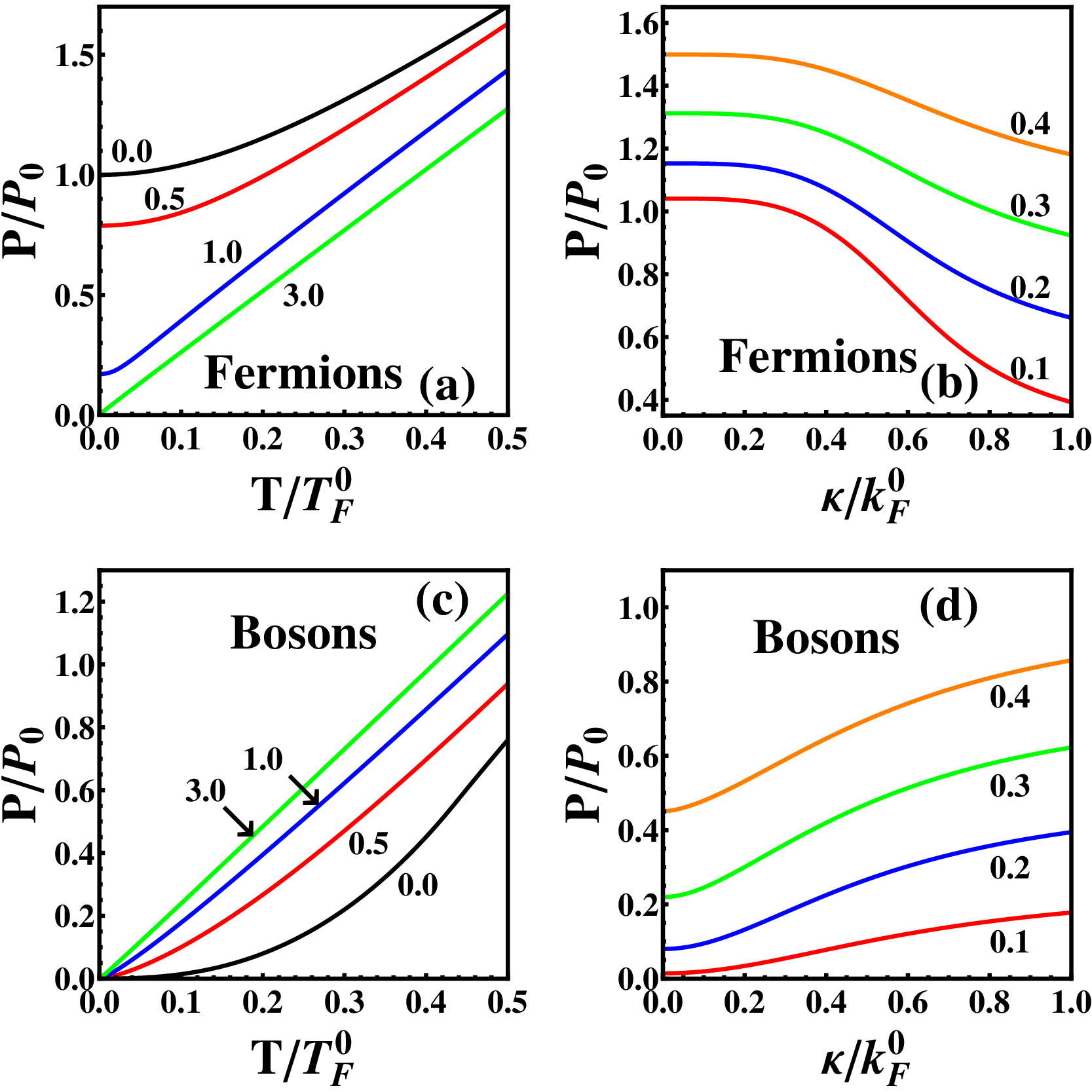}
\parbox{130mm}{\caption{\label{fig-pressure}\small
(Color online). Variation of rescaled pressure with $\tilde T$ for four values of $\tilde\kappa$ in (a) and (c), and with $\tilde\kappa$ for four values of $\tilde T$ in (b) and (d).}}
\end{center}
\begin{center}
\includegraphics[width=80mm,angle=0]{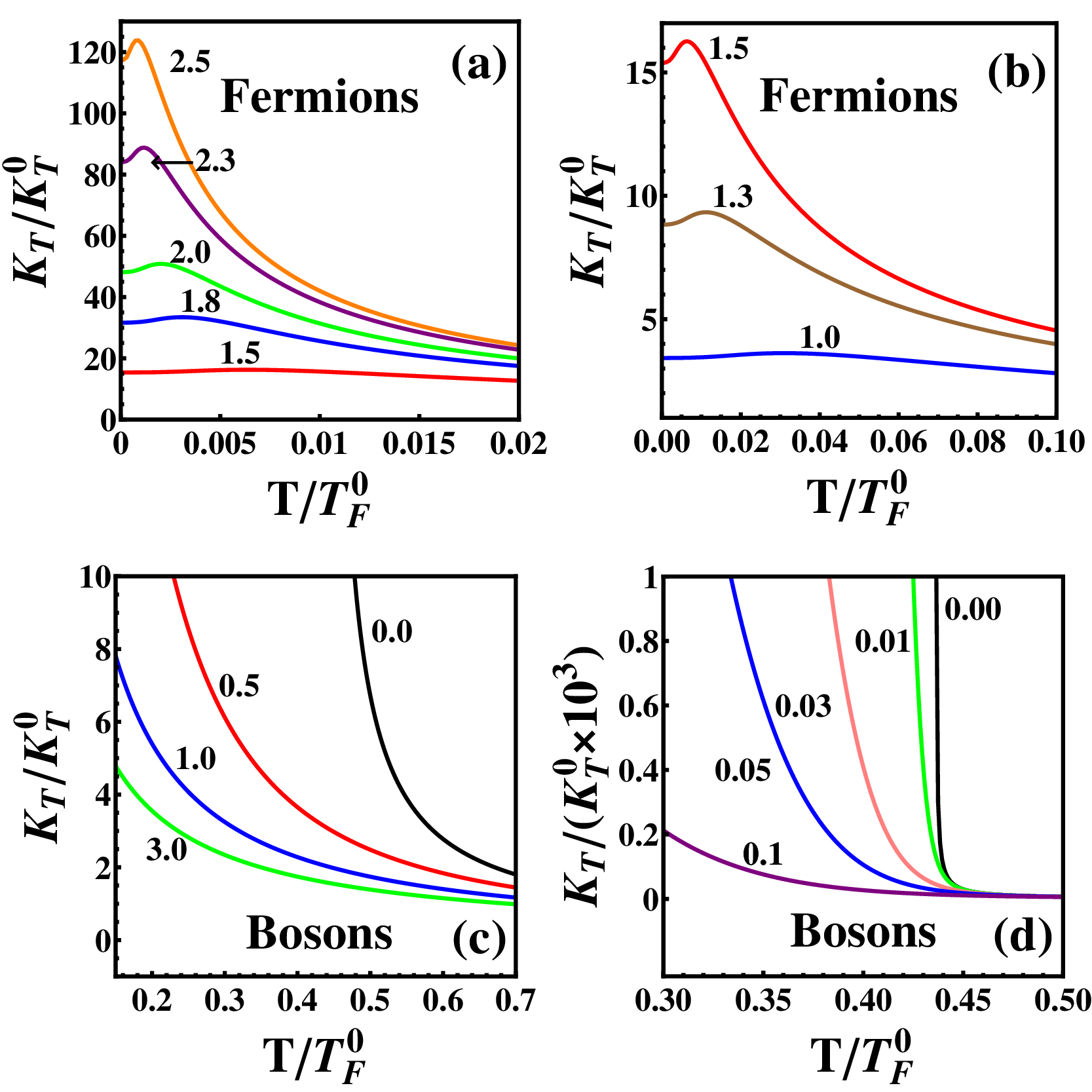}
\parbox{130mm}{\caption{\label{fig-compress}\small
(Color online). Rescaled compressibility as a function of $\tilde T$ for various values of $\tilde\kappa$ labeled therein.}}
\end{center}
\end{figure}
\noindent\textbf{B. Harmonic Trapping}

The harmonically trapped thermodynamic systems are characterized by the extensive variable, ``harmonic" volume ${\cal{V}}=1/\omega^3$, and the conjugate intensive variable, ``harmonic" pressure ${\cal{P}}=-\left(\partial\Omega_H/\partial{\cal{V}}\right)_{\mu,T}$. The semiclassical expression for the grand potential is obtained using Eq.(\ref{dos-harmonic}) together with Eq.(\ref{polylog}) in
$\Omega_H(\mu,T,{\cal{V}})=(1/\eta\beta)\int_0^\infty G_H(\varepsilon)\ln\left(1-\eta z\e^{-\beta\varepsilon}\right)\de\varepsilon$ and is given by
\1
\label{omega-H}
\Omega_H=-\frac{2{\cal{V}}}{\eta\hbar^3\beta^4}\left[\rm{Li}_{4}\left(\eta z\right)+\phi\left(\kappa\Lambda\right)\rm{Li}_{3}
\left(\eta z\right)\right].
\2
The number of particles and the energy are now given by
\1
\label{number-H}
N=\frac{2{\cal{V}}}{\eta\hbar^3\beta^3}\left[\rm{Li}_{3}\left(\eta z\right)+\phi\left(\kappa\Lambda\right)\rm{Li}_{2}\left(\eta z\right)\right]
\2
\1
\label{energy-H}
E=\frac{2{\cal{V}}}{\eta\hbar^3\beta^4}\left[3\rm{Li}_{4}\left(\eta z\right)+2\phi\left(\kappa\Lambda\right)\rm{Li}_{3}\left(\eta z\right)\right]
\2
and the EOS can be written in the form
\1
\label{EOS-harmonic}
{\cal{P}}{\cal{V}}=\frac{1}{3}\left[E+\frac{1}{4}\left(\frac{\partial E}{\partial\ln\kappa}\right)_{\mu}\right].
\2
 This gives ${\cal{P}}{\cal{V}}=\frac{1}{3}E$ for $\kappa\to 0$ as the EOS of a harmonically trapped gas without SOC is just like that of ultrarelativistic or massless particles \cite{romero1:prl05}.

The expressions for the other quantities are now
\ea
\label{pressure-H}
{\cal{P}}=\frac{2}{\eta\hbar^3\beta^4}\left[\rm{Li}_{4}\left(\eta z\right)+\phi\left(\kappa\Lambda\right)\rm{Li}_{3}
\left(\eta z\right)\right],
\ee
\1
\label{free energy-H}
F =\mu N- \frac{2{\cal{V}}}{\eta\hbar^3\beta^4}\left[\rm{Li}_{4}\left(\eta z\right)+2\phi\left(\kappa\Lambda\right)\rm{Li}_{3}\left(\eta z\right)\right],
\2
\1
\label{entropy-H}
\frac{S}{N k_B} =\frac{4\rm{Li}_{4}\left(\eta z\right)+3\phi\left(\kappa\Lambda\right)
\rm{Li}_{3}\left(\eta z\right)}{\rm{Li}_{3}\left(\eta z\right)+\phi\left(\kappa\Lambda\right)
\rm{Li}_{2}\left(\eta z\right)}-\beta\mu,
\2
\1
\label{compress-H}
K_T=\left(\frac{\beta}{n}\right)\frac{\rm{Li}_{2}\left(\eta z\right)+\phi\left(\kappa\Lambda\right)
\rm{Li}_{1}\left(\eta z\right)}{\rm{Li}_{3}\left(\eta z\right)+\phi\left(\kappa\Lambda\right)
\rm{Li}_{2}\left(\eta z\right)}
\2
\ea
\frac{C_{\cal{V}}}{Nk_{\mathrm{B}}}&=&\frac{6\left[2\rm{Li}_{4}\left(\eta z\right)+\phi\left(\kappa\Lambda\right)
\rm{Li}_{3}\left(\eta z\right)\right]}{\rm{Li}_{3}\left(\eta z\right)+\phi\left(\kappa\Lambda\right)\rm{Li}_{2}\left(\eta z\right)}\nonumber\\
&&-\frac{\left[3\rm{Li}_{3}\left(\eta z\right)+2\phi\left(\kappa\Lambda\right)
\rm{Li}_{2}\left(\eta z\right)\right]^2}{\left[\rm{Li}_{3}\left(\eta z\right)+\phi\left(\kappa\Lambda\right)\rm{Li}_{2}\left(\eta z\right)\right] \left[\rm{Li}_{2}\left(\eta z\right)+\phi\left(\kappa\Lambda\right)\rm{Li}_{1}\left(\eta z\right)\right]}
\ee
and
\ea
T\alpha_{\mathrm{T}}&=&\frac{4\left[\rm{Li}_{4}\left(\eta z\right)+\phi\left(\kappa\Lambda\right)
\rm{Li}_{3}\left(\eta z\right)\right]\left[\rm{Li}_{2}\left(\eta z\right)+\phi\left(\kappa\Lambda\right)\rm{Li}_{1}\left(\eta z\right)
\right]}{\left[\rm{Li}_{3}\left(\eta z\right)+\phi\left(\kappa\Lambda\right)\rm{Li}_{2}\left(\eta z\right)\right]^2}\nonumber\\
&&-\phi\left(\kappa\Lambda\right)\rm{Li}_{3}\left(\eta z\right)\frac{\left[\rm{Li}_{2}\left(\eta z\right)+\phi\left(\kappa\Lambda\right)
\rm{Li}_{1}\left(\eta z\right)\right]}{\left[\rm{Li}_{3}\left(\eta z\right)+\phi\left(\kappa\Lambda\right)\rm{Li}_{2}\left(\eta z\right)\right]^2}\nonumber\\
&&-\frac{\left[3\rm{Li}_{3}\left(\eta z\right)+2\phi\left(\kappa\Lambda\right)
\rm{Li}_{2}\left(\eta z\right)\right]}{\left[\rm{Li}_{3}\left(\eta z\right)+\phi\left(\kappa\Lambda\right)\rm{Li}_{2}\left(\eta z\right)\right]}
\ee

All the expressions in this subsection are new and give, under the limit $\kappa\to 0$, the results for the harmonically trapped ideal gases without SOC; the corresponding Bose gas results \cite{romero1:prl05} are thus retrieved keeping in mind that here we are dealing with pseudospin-$\frac{1}{2}$ bosons. Furthermore, the thermodynamic relations involving $C_{\cal{P}}, C_{\cal{V}}, \alpha_{\mathrm{T}}, K_{\mathrm{T}}$ and $K_{\mathrm{S}}$ mentioned in the previous subsection become helpful here too.

\pagebreak
\noindent\textbf{IV. NUMERICAL RESULTS AND DISCUSSIONS}

The variation of  rescaled pressure $\tilde P=P/P_0$ with $\tilde {T}$ (or $\tilde{\kappa}$) for various fixed values of $\tilde{\kappa}$ (or $\tilde {T}$) is shown in Figs. \ref{fig-pressure}(a) to \ref{fig-pressure}(d). Here $P_0=\left(2n\varepsilon^0_F/5\right)$ is the pressure of spin-$\frac{1}{2}$ fermions without SOC at $T=0$ and $n$ is the number density.  It can be seen from plots that pressure decreases for fermions but increases for bosons with increase in the Weyl coupling at any fixed temperature. In order to have a closer look at this behavior analytically, we proceed to construct the virial expansion for the EOS in the form
\1
\label{virial}
P\beta=\sum_{\ell=1}^\infty b_{\ell} (T, \kappa)n^{\ell},
\2
where $b_{\ell}(T,\kappa)$  is the $\ell$th virial coefficient with $b_1(T, \kappa)=1$.

We obtain the expansion for the fugacity, from inversion of power-series in Eq. (\ref{number-series}), and write explicitly only up to third-order terms in a  parameter
\1
\label{xi-defined}
\xi=\frac{n\Lambda^3}{2\left(1+\phi\left(\kappa\Lambda\right)\right)}.
\2
We then get, on writing $\phi\equiv\phi\left(\kappa\Lambda\right)$,
\ea
\label{z-expansion}
z= \xi\left[1-\frac{\eta\xi}{2^{3/2}} \frac{1+2\phi}{1+\phi} + \frac{\eta^2\xi^2}{2^2}
\left\{\left(\frac{1+2\phi}{1+\phi}\right)^2-\frac{2^2}{3^{3/2}}
\frac{1+3\phi}{1+\phi}\right\}-\eta^3 {\cal O}\left(\xi^3\right)\right].
\ee
Using Eqs. (\ref{number-series}) and (\ref{pressure-series}), one has
\ea
\label{pressure-series2}
P\beta=n\left[1-\frac{\eta z}{2^{5/2}} \frac{1+2\phi }{1+\phi }+\frac{\eta^2 z^2}{2^4}
\left\{\left(\frac{1+2\phi }{1+\phi }\right)^2-\frac{2^5}{3^{5/2}}
 \frac{1+3\phi}{1+\phi } \right\} -\eta^3 {\cal O}\left(z^3\right)\right].
\ee
Combination of Eqs. (\ref{z-expansion}) and (\ref{pressure-series2}) yield finally
\ea
\label{virial-truncated}
P\beta=n\left[1-\frac{\eta\xi}{2^{5/2}} \frac{1+2\phi }{1+\phi }+\frac{\eta^2\xi^2}{2^3}
\left\{\left(\frac{1+2\phi }{1+\phi }\right)^2-\frac{2^4}{3^{5/2}}
\frac{1+3\phi }{1+\phi }\right\}
-\eta^3 {\cal O}\left(\xi^3\right)\right].
\ee
which reduces, on putting $\kappa=0$, to the textbook results \cite{lal:93, pab:11} in powers of $\xi=n\Lambda^3/2$ for spin-$\frac{1}{2}$ particles.

\begin{figure}
\begin{center}
\includegraphics[width=80mm,angle=0]{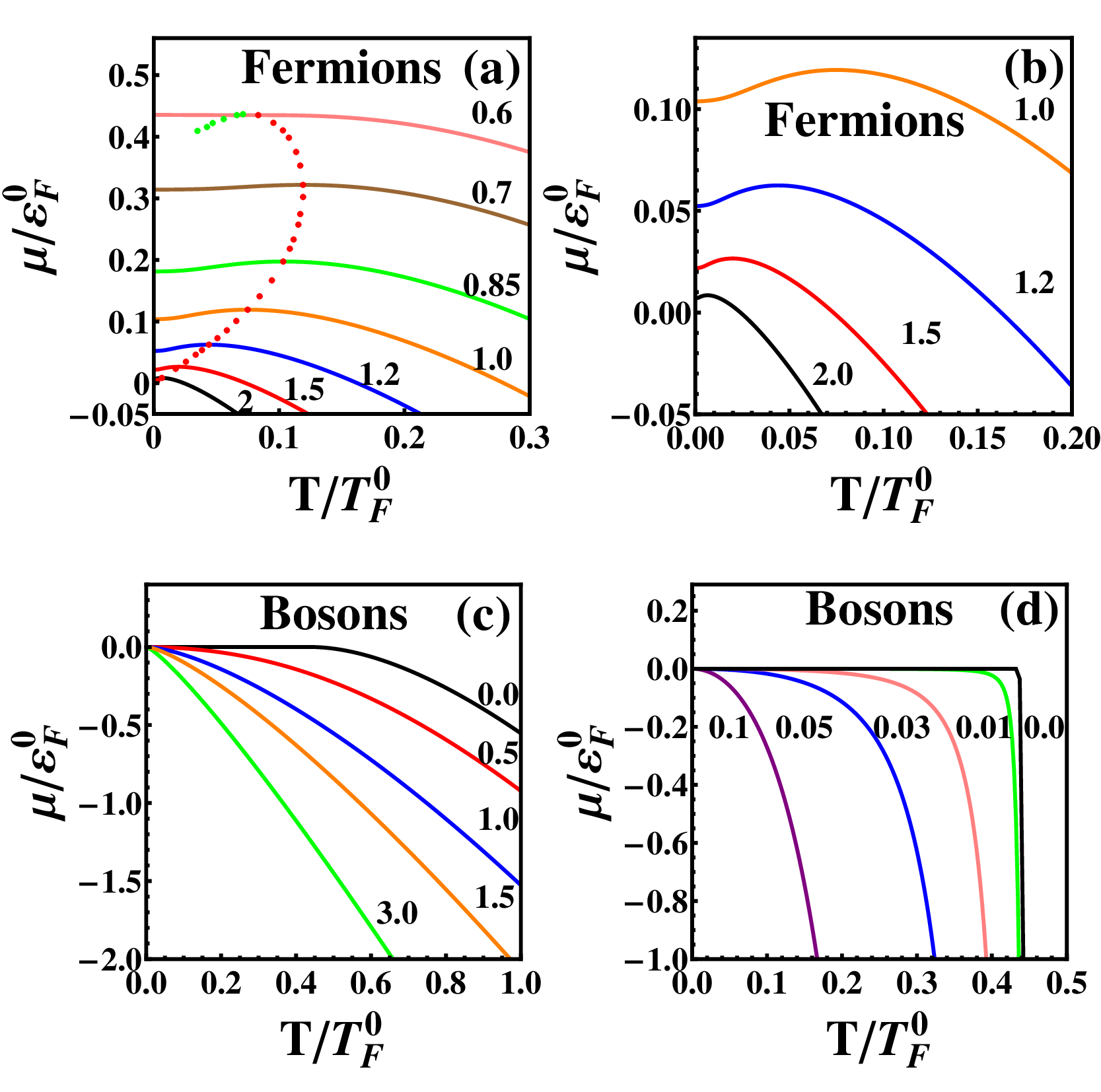}
\parbox{130mm}{\caption{\label{fig-chempot}\small
(Color online). Variation of rescaled chemical potential as a function of $\tilde T$ for various values of $\tilde\kappa$. Dotted curve in (a) is the locus of finite-$T$ extrema; starting from peak location, clockwise is for maxima (red dots) and anticlockwise for minima (green dots).}}
\end{center}
\begin{center}
\includegraphics[width=80mm,angle=0]{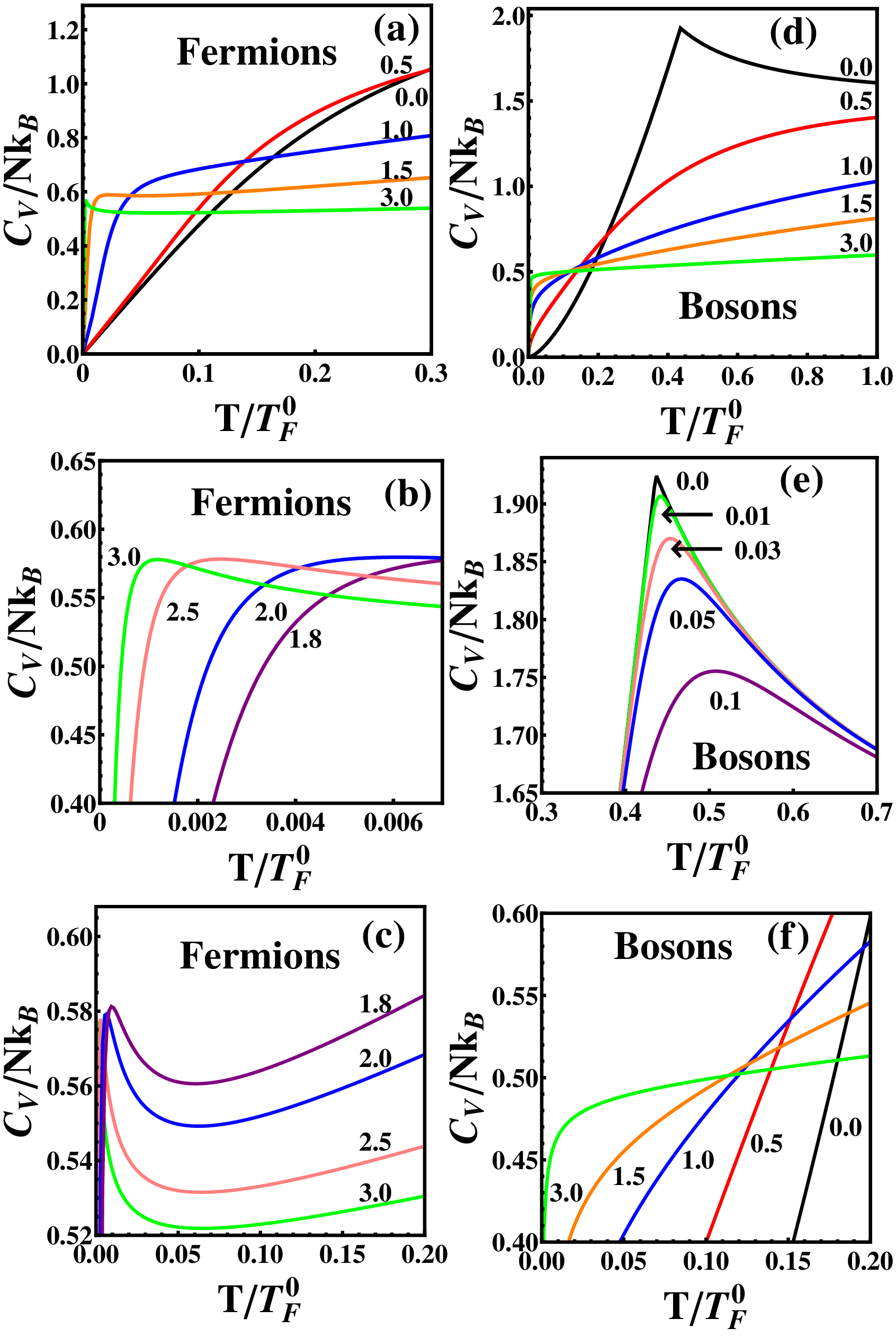}
\parbox{130mm}{\caption{\label{fig-spheat}\small
(Color online). Variation of rescaled specific heat  as a function of $\tilde T$ for various values of $\tilde\kappa$.}}
\end{center}
\end{figure}

Let us substitute into Eq. (\ref{virial-truncated}) the expression for $\xi$ given in Eq. (\ref{xi-defined}) and compare the resulting equation with Eq. (\ref{virial}). This leads to the inequality $b_2\left(T, \kappa\right)/b_2\left(T, \kappa\g0\right)=(1+2\phi)/(1+\phi)^2\leq1$ implying that $b_2(T, \kappa)$ becomes less negative (positive) for bosons (fermions) as soon as the SO coupling is switched on. Since the second virial coefficient represents two-particle interactions \cite{lal:93}, it  can be interpreted that \emph{the Weyl coupling induces} \emph{repulsive} (\emph{attractive}) \emph{statistical interactions in bosons} (\emph{fermions}).  Hence the effect of the SOC is to weaken the ``effective" interactions  resulting into the behavior of the pressure as depicted in Figs. \ref{fig-pressure} (a)--(d).
Furthermore, the decrease (increase) in the pressure suggests that the SOC makes the gas  more (less) compressible which is supported by general trends in Figs. \ref{fig-compress}(a) to \ref{fig-compress}(d) wherein rescaled isothermal compressibility $\tilde K_\mathrm{T}=K_\mathrm{T}/K^0_\mathrm{T}$ has been plotted as a function of $\tilde\kappa$ at several temperatures with $K^0_\mathrm{T}=3/(2n\varepsilon_F^0)$ as isothermal compressibility of an ideal Fermi gas without SOC at $T=0$. The decrease in the Pauli pressure and consequent increase in the isothermal compressibility due to increase in Rashba coupling strength had been found by Han and S$\acute{a}$ de Melo \cite{has:pra12} in course of their study of the BCS-BEC crossover in a balanced mixture of 3D ultracold fermions. The decrease in Pauli pressure has been found by Ghosh et al. \cite{ghosh3:pra11} also.

The bosonic curves for very low $\kappa$ in \ref{fig-compress}(d) are showing the artifact of diverging $K_T$ present in an ideal Bose gas for $T<T^0_{\mathrm{C}}\simeq 0.436 T^0_{\mathrm{F}}$. On the other hand, a fermionic curve has a peak at very low temperature which becomes more and more pronounced and shifts towards lower $T$ as $\kappa$ increases (cf. Figs.\ref{fig-compress} (a) and (b)). Since $K_\mathrm{T}$ is related to the derivative of $\mu$, any anomalous behavior in one could be expected to have its signature in the other. Hence we plot in Figs. \ref{fig-chempot} (a)--(d) the chemical potential $\mu(T,\kappa)$ as a function of $T$ for various values of $\kappa$ calculated from the implicit relation given by Eq. (\ref{number}). These computed values have, in fact, gone as input in calculations of all the thermodynamic quantities displayed in Figs. \ref{fig-pressure}, \ref{fig-compress} and \ref{fig-spheat}. Figures \ref{fig-chempot}(b) and \ref{fig-chempot}(d) depict magnified graphs for some small ranges of $T$ to reveal the structures appearing therein.

It can be seen in Fig. \ref{fig-chempot}(d) that the nature of the curves reflect incipient Bose-Einstein condensation for very low values of $\kappa$, just like the plots for $K_{\mathrm{T}}$ displayed in \ref{fig-compress}(d), although the Bose-condensation is inhibited by the presence of coupling-weighted 1D character in the DOS (cf. Eq. (\ref{dos})).
For fermions, the numerical values of $\mu(T, \kappa)$ reveal interesting behavior in low-$T$ regime. For small $\kappa$, $\mu(T, \kappa)$ is increasing with decrease in $T$ just like in the case of an ideal 3D Fermi gas without SOC. However, starting at $\kappa\approx0.5\,k_F^0$, the fermion chemical potential exhibits a shoulder near $T\approx0.12\,T_F^0$ which grows upon further increase of $\kappa$ and eventually forms one local maximum and one local minimum whose loci are being depicted in Fig. \ref{fig-chempot}(a) by the dotted curve. The right portion (red dots) of the peak of this curve is the locus of maxima whereas the left portion (green dots) is only a part of the locus of minima, it touches the vertical axis at $\tilde\mu\approx0.40$ when $\kappa=\kappa_{\mathrm{c}}\approx0.63 k^0_F$ as described in Appendix \ref{appen-chempot} where the detailed analysis of $\mu(T,\kappa)$ versus $T$ for successively increasing $\kappa$ starting from zero has been presented.

It is worth mentioning that Ref. \cite{gds:epjd03} has discussed temperature variation of the chemical potential for an ideal Fermi gas without SOC wherein a low-$T$ peak has been found in less than 2D. Also, it  has been discussed in \cite{sas:arxiv13} that an ideal Fermi gas confined between penetrable multilayers or multitubes has effective reduced dimensionality and there is dimensional crossovers from 3D to 2D and then to 1D as the impenetrability increases resulting into appearance of peak in $\mu$ as a function of $T$.
Hence the presence of both extrema in a curve in our work cannot be understood solely in terms of SO-coupling induced dimensionality reduction and its probable explanation is given in Appendix C. Also, it is clear from Eq.(\ref{approx-mu}) that fermionic $\mu(T,\kappa)$  has different behavior depending on whether the Fermi level $\varepsilon_F(\kappa)\equiv\mu(T=0,\kappa)$ of a SO-coupled system is above or below $\varepsilon_{\mathrm{so}}$.

Figures \ref{fig-spheat}(a)-(f) show temperature-dependence of $C_V$ for various values of $\kappa$. We have from Figs. \ref{fig-spheat}(a) and \ref{fig-spheat}(d) that $(C_V/N k_B)\to3/2$ ($1/2$) at high temperatures consistent with the Dulong-Petit law for low-$\kappa$ (high- $\kappa$) when the system is expected to have quasi-3D (quasi-1D) behavior. Also, for $T\to0, (C_V/N k_B)\to0$ as per the third law of thermodynamics. Figures \ref{fig-spheat}(b) and \ref{fig-spheat}(c) are blown up curves in low-$T$ regions to depict clearly the structures in the curves. It is found from  Fig. \ref{fig-spheat}(c) that as $T$ decreases, each curve has a dip before showing a peak at very low temperature, which is again an anomalous behavior. This kind of structure in the specific heat curve for fermions in reduced effective dimension has been reported in Ref. \cite{sas:arxiv13} also. Furthermore, it can be seen that Fig. \ref{fig-spheat}(e) supports SO-coupled incipient Bose-condensation, and crossings of the curves for various $\kappa$ in Figs. \ref{fig-spheat}(b) and \ref{fig-spheat}(f) are required by the Dulong-Petit law since the effective dimensionality changes as a consequence of change in the coupling strength.
\begin{figure}
\begin{center}
\includegraphics[width=80mm,angle=0]{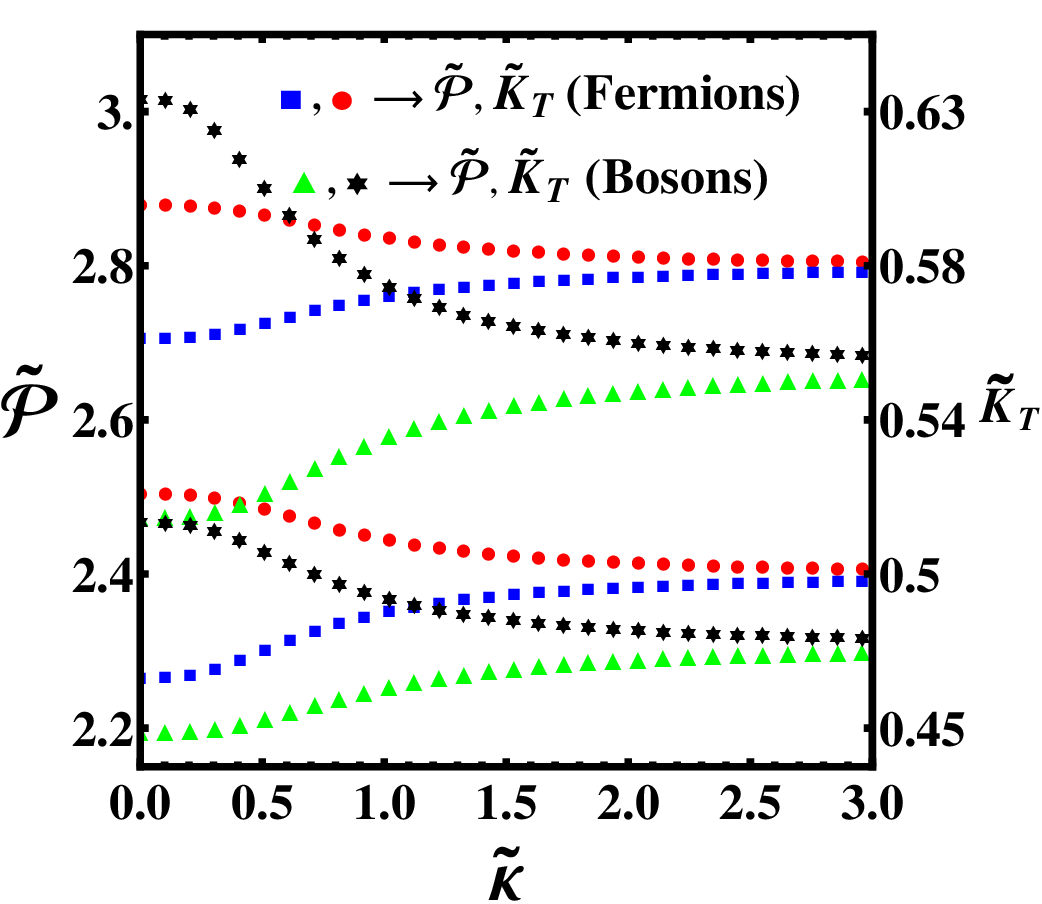}
\parbox{130mm}{\caption{\label{fig-trappedpressure}\small
(Color online). $\tilde{\cal{P}}-\tilde{\kappa}$ and $\tilde{K_{\mathrm{T}}}-\tilde{\kappa}$ isotherms for harmonically trapped gases. Set of lower (upper) four curves are for $\tilde{T}=0.6$ ($0.7$). Left (right) vertical axis has values of $\tilde{\cal{P}}$ ($\tilde{K_{\mathrm{T}}}$). Symbols triangle and square (star and circle) are plots for $\tilde{\cal{P}}$ ($\tilde{K_{\mathrm{T}}}$). Also, triangle and star (square and circle) represent bosons (fermions).}}
\end{center}
\begin{center}
\includegraphics[width=100mm,angle=0]{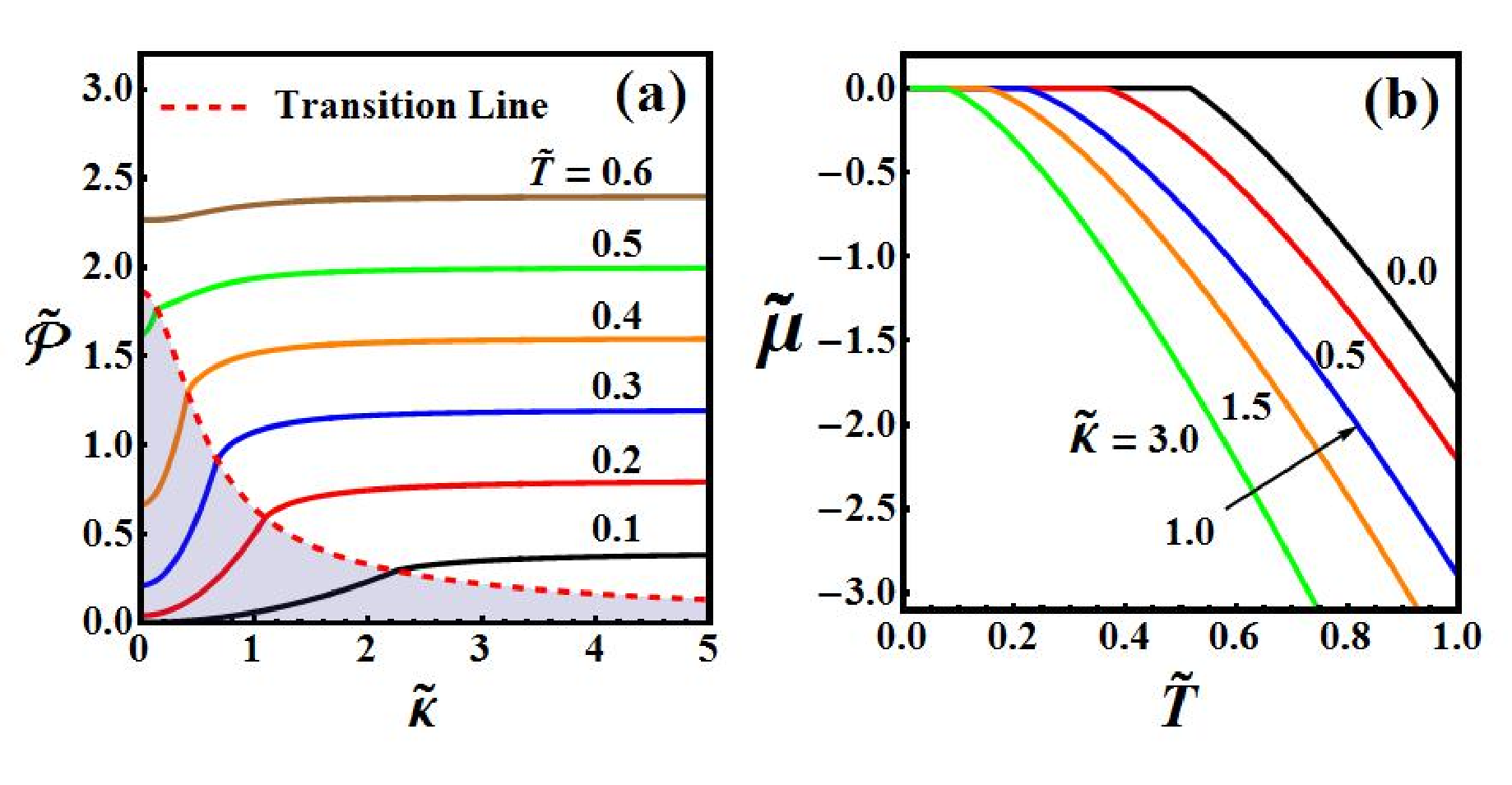}
\parbox{130mm}{\caption{\label{fig-condensation}\small
(Color online). Bosonic phase diagrams (a) $\tilde{\cal{P}}-\kappa$ and (b) $\tilde{\mu}-T$ for values of $\tilde{T}$ and $\tilde{\kappa}$ labeling the curves. Shaded region represents the Bose-condensed phase.}}
\end{center}
\end{figure}

The isotherms ${\cal{P}}-\kappa$ and $K_{\mathrm{T}}-\kappa$ are shown combinedly in Fig. 5 for harmonically trapped noninteracting bosons and fermions; the values of the rescaled quantities $\tilde{{\cal{P}}}={\cal{P}}(\kappa)/{\cal{P}}(0)$ and $\tilde{K_{\mathrm{T}}}=K_{\mathrm{T}}(\kappa)/K_{\mathrm{T}}(0)$ are on left and right vertical axes, respectively. We have introduced the symbol for any physical quantity $\chi(\kappa=0)$ as simply  $\chi(0)$. The lower set of four curves starting from the lowermost, two dashed (fermions, bosons) and two solid (bosons, fermions), is for the reduced temperature $\tilde{T}=T/T_\mathrm{F}(0)=0.6$ and the upper set for $\tilde{T}=0.7$. The reduced coupling strength is $\tilde{\kappa}=\kappa/k_{\mathrm{F}}(0)$. An SO-coupled trapped Bose gas shows Bose-Einstein condensation, unlike SO-coupled uniform gas, and hence temperatures have been chosen to correspond to the Bose gas in the normal phase for any $\kappa$. Also, ${\cal{P}}(0)=\varepsilon^4_{\mathrm{F}}(0)/12\hbar^3=n\varepsilon_{\mathrm{F}}(0)/4$ and $K_{\mathrm{T}}(0)=3/n\varepsilon_{\mathrm{F}}(0)$ with $\varepsilon_{\mathrm{F}}(0)$ as the Fermi energy for harmonically trapped fermions without SOC. The graphs in  Fig. 5 suggest that the conclusion regarding the SOC-induced interactions is the same here as that described earlier for uniform gases, i.e. \emph{the increase in the coupling strength makes the statistical interaction weaker and weaker in trapped gases too}.

Figure 6(a) shows bosonic ${\cal{P}}-\kappa$ phase diagram wherein each curve is labeled by its value of $\tilde{T}$, the dashed curve is the transition line for the Bose-Einstein condensation, and the shaded region represents the condensed phase. It can be seen that increase in the SOC strength decreases the transition temperature which is corroborated by the $\mu-T$ phase diagram, with $\tilde{\mu}=\mu/\varepsilon_{\mathrm{F}}(0)$ for various values of $\tilde{\kappa}$ depicted in 6(b). This kind of $T_{\mathrm{C}}$ versus $\kappa$ behavior is consistent with the weakening effect of the statistical attractive interaction on increasing the value of $\kappa$. However, it is at variance with the study of Hu and Liu \cite{hul:pra12} who observed that in the thermodynamic limit,  $T_{\mathrm{C}}$ would not get affected by the Rashba coupling. Our study is for the Weyl coupling and separate study for the Rashba coupling might be required to resolve the issue that the latter would have zero effective statistical interaction in trapped bosons whereas it does have finite effective interaction in uniform Fermi gases as mentioned earlier in this section while analyzing $T=0$ results of Han and S$\acute{a}$ de Melo \cite{has:pra12}, and of Ghosh \emph{et al}. \cite{ghosh3:pra11}.

\noindent\textbf{V. SUMMARY AND OUTLOOK}

We have developed exact analytical expressions for the grand potential in uniform as well as harmonically trapped cases in a unified form for noninteracting Bose, Fermi and Boltzmann gases having 3D isotropic spin-orbit coupling, called the Weyl coupling. The expressions for various thermodynamic quantities have been derived thereafter, and their computed values have been depicted graphically as a function of temperature as well as the spin-orbit coupling constant. It is found that the Weyl coupling induces statistical interactions which weaken the effective interparticle interactions present in the systems without spin-orbit coupling: bosons have reduced effective attraction and fermions have reduced effective repulsion whether the gas is uniform or trapped. The conclusion has been arrived at from graphical depiction of pressure in both the situations; the result has been corroborated by plots of the isothermal compressibility and analytically by the virial expansion of the EOS in the uniform system. The variation of chemical potential and specific heat have also been analyzed and anomalous behavior revealed in case of uniform gases. The plots of internal energy, free energy and entropy have not been displayed here as they do not show any worth describing feature.

Our studies suggest that the physics due to synthetic spin-orbit coupling is richer than previously appreciated -- not only the theoretically investigated features explained plausibly in terms of coupling-induced enhanced DOS but some additional features also can be understood in terms of coupling-induced modifications in interparticle statistical interactions. It is found that a uniform Bose gas with very weak coupling shows incipient Bose-Einstein condensation and a Fermi system has different behaviors on varying the coupling constant if the Fermi level is above or below the Dirac point. Our work for uniform gases in conjunction with that in Ref. \cite{hey:pra11} would constitute complete thermodynamics of noninteracting atomic quantum gases with isotropic 3D spin-orbit coupling. Most of the expressions derived for the thermodynamic quantities of uniform gases and all those of harmonically trapped gases are new, and the study of complete thermodynamics would be possible also for harmonically trapped gases with Weyl SOC  in any range of physical parameters. Moreover, we have pointed out that the results obtained in Refs. \cite{hul:pra12} and \cite{has:pra12} for Rashba coupling are at variance with each other if analyzed in terms of effective statistical interactions.

The fluctuation-dissipation theorem gives the relation \cite{pis:03} $\lim_{q\to0}S(q,T)=n k_B T K_T$ with $S(q,T)$ as the temperature-dependent static structure factor.  Hence the enhanced (diminished) compressibility of the gas, along with any structure therein, would get reflected in measurement of enhanced (diminished) long-wavelength fluctuation of the bunching (antibunching) effect \cite{bps:pre11, brs:prl13}, quantified by $S(q\approx0,T)$, in bosons (fermions). Our theoretical study of $S(q,T)$ and hence the pair correlation function leading to the statistical interparticle interaction potential will form part of a separate communication.

\noindent\textbf{ACKNOWLEDGMENTS}\\
RG acknowledges grant of research fellowship by the MHRD,  and thanks Zeng-Qiang Yu for useful correspondence.\\

\noindent $^{\dag}$Corresponding author: gss.phy@iitr.ac.in\\

\section*{Appendix}

\renewcommand{\theequation}{\mbox{\Alph{section}.\arabic{equation}}}
\begin{appendix}

\setcounter{equation}{0}
\section{Density of states for SO coupled uniform gases}
\label{appen-dos}
The use of dispersion relations given in Eq. (\ref{dispersion}) gives the forms of partial DOS, $\rho_{\pm}\left(\varepsilon\right)\equiv\rho_{\pm}\left(\varepsilon, \kappa\right)=(1/V)\sum_{\textbf{k}}\delta\left(\varepsilon - \varepsilon_{\pm}\left(\textbf{k}\right)\right)$, as
\1
\label{rho-plus}
\rho_{+}\left(\varepsilon\right)=\frac{1}{2}~\rho^{0}_{3D}\left(\varepsilon\right)\left(1-\sqrt{\frac{\varepsilon_{\rm{so}}}{\varepsilon}}~\right)^2
\Theta\left(\varepsilon-\varepsilon_{\rm{so}}\right)
\2
and
\1
\label{rho-minus}
\rho_{-}\left(\varepsilon\right)=\rho^{0}_{3D}\left(\varepsilon\right)\left[\left(1+\frac{\varepsilon_{\rm{so}}}{\varepsilon}\right)
\Theta\left(\varepsilon_{\rm{so}}-\varepsilon\right)+
\frac{1}{2}\left(1+\sqrt{\frac{\varepsilon_{\rm{so}}}{\varepsilon}}~\right)^2
\Theta\left(\varepsilon-\varepsilon_{\rm{so}}\right)\right]
\2
whence we have the expression for $\rho(\varepsilon)$ as given in Eq. (\ref{dos}).

\begin{figure}
\begin{center}
\includegraphics[width=80mm,angle=0]{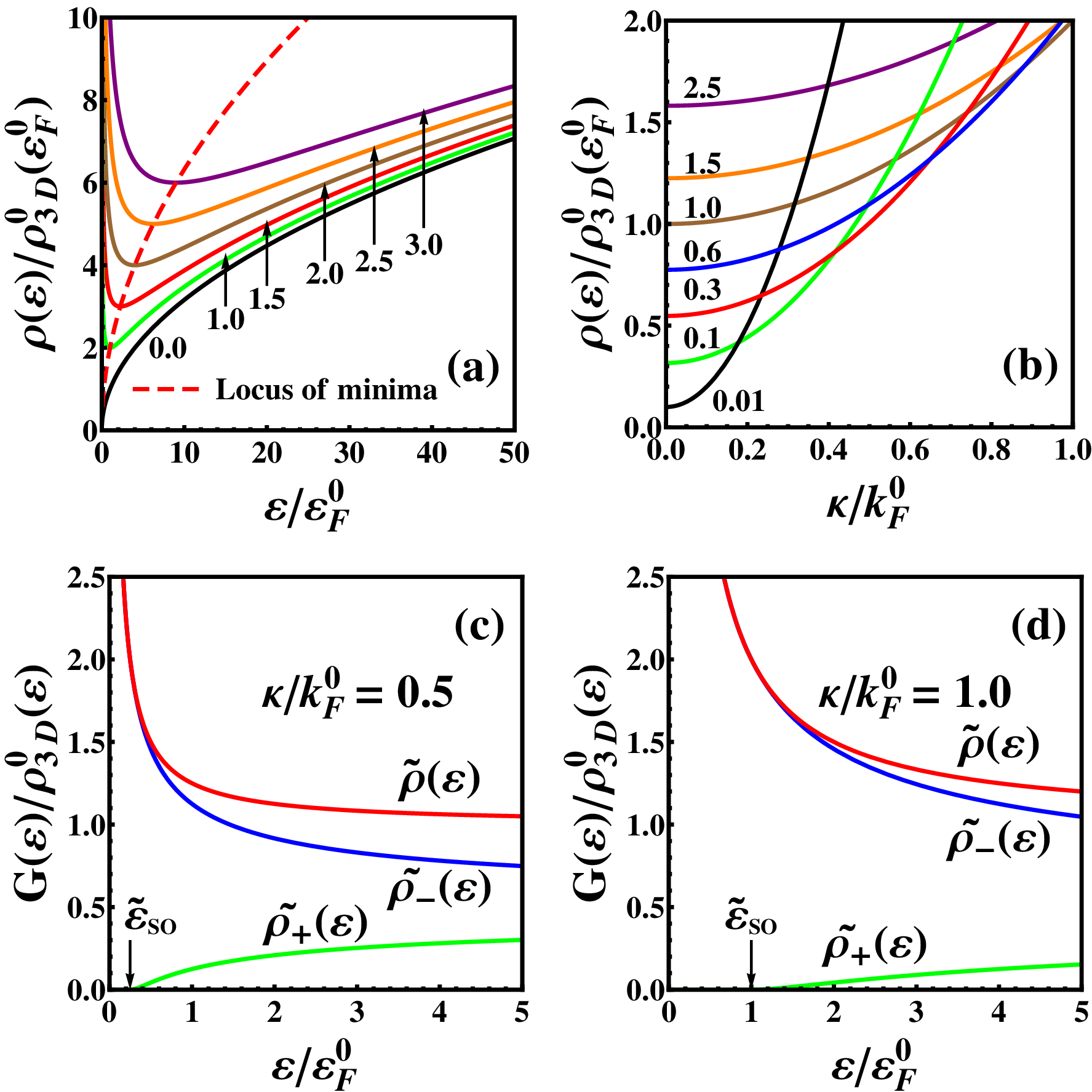}
\parbox{140mm}{\caption{\label{fig-dos4}\small
(Color online). Variation of rescaled DOS with $\tilde{\varepsilon}$ in (a), (c) and (d), and with $\tilde\kappa$ in (b). Values of $\tilde\kappa$ are labeled in (a), and $\tilde\varepsilon$ in (b). $G(\varepsilon)$ in (c) and (d) stands for $\rho_{+}\left(\varepsilon\right)$, $\rho_{-}\left(\varepsilon\right)$ or $\rho\left(\varepsilon\right)$. Location of $\tilde{\varepsilon}$ for minimum in $\rho\left(\varepsilon\right)$ is marked by arrow below which the DOS for + helicity branch is zero.}}
\end{center}
\end{figure}

It is found that $\rho\left(\varepsilon\right)$ has the minimum value $\rho_{\min}\left(\varepsilon\right) = 2\rho^{0}_{3D}\left(\varepsilon_{\rm{so}}\right)$ at $\varepsilon=\varepsilon_{\rm{so}}$. The minimum arises because $\rho_-\left(\varepsilon\right)$ decreases continuously in the range $0\le\varepsilon\le\infty$ but $\rho_+\left(\varepsilon\right)\g0$ for $\varepsilon<\varepsilon_{\rm{so}}$ and increases monotonically for $\varepsilon>\varepsilon_{\rm{so}}$. The variation of rescaled DOS, $\tilde{\rho}\left(\varepsilon\right)= \rho\left(\varepsilon\right)/\rho^{0}_{3D}\left(\varepsilon^0_F\right)$, as a function of reduced energy $\tilde\varepsilon=\varepsilon/\varepsilon_F^0$ is depicted in Fig. \ref{fig-dos4}(a) wherein dashed curve is the locus of minima in the DOS for various values of $\tilde\kappa$. The plots of rescaled $\rho_+\left(\varepsilon\right)$, $\rho_-\left(\varepsilon\right)$ and $\rho\left(\varepsilon\right)$ with respect to $\rho^0_{3D}\left(\varepsilon\right)$ as a function of  $\tilde{\varepsilon}$ for $\tilde\kappa=0.5$ and $1.0$ are shown respectively in Figs. \ref{fig-dos4}(c) and \ref{fig-dos4}(d). Whereas the value of $\rho_+\left(\varepsilon\right)$ goes down for increased $\kappa$ and a given value of $\varepsilon$, the corresponding values of both $\rho_-\left(\varepsilon\right)$ and $\rho\left(\varepsilon\right)$ go up. Also, it is seen in Fig. \ref{fig-dos4}(b) that $\rho\left(\varepsilon\right)$ versus $\kappa$ curve is parabolic which gets flattened for an increased value of energy.

\setcounter{equation}{0}
\section{Isochoric specific heat for SO coupled uniform gases}
\label{appen-spheat}
The relation $C_V=\left(\partial E/\partial T\right)_{V,N}$ gives, on using Eq.(\ref{energy}) and the recurrence relation (\ref{recurrence}),
\ea
\label{spheat1}
C_V=\frac{3}{2}\frac{k_BV}{\eta\Lambda^3}\left[5\rm{Li}_{5/2}\left(\eta z\right)+\phi\left(\kappa\Lambda\right)\rm{Li}_{3/2}\left(\eta z\right)\right]~~~~~~~~~~~~~~\nonumber\\
+\frac{V}{\eta\beta\Lambda^3}\left[3\rm{Li}_{3/2}\left(\eta z\right)+\phi\left(\kappa\Lambda\right)\rm{Li}_{1/2}\left(\eta z\right)\right]
\frac{1}{z}\left(\frac{\partial z}{\partial T}\right)_{V,N}.
\ee
In order to find out an expression for $\left(1/z\right)\left(\partial z/\partial T\right)_{V,N}$, we rewrite Eq. (\ref{number}) as
\1
\label{number-rewrite}
\rm{Li}_{3/2}\left(\eta z\right)+\phi\left(\kappa\Lambda\right)\rm{Li}_{1/2}\left(\eta z\right)=N\eta\Lambda^3/2V
\2
and apply the operator $(\partial/\partial T)_{V,N}$ from left on both the sides. We then use in the resulting expression Eq. (\ref{recurrence}) on the left hand side and Eq. (\ref{number-rewrite}) on the right hand side. The process gives finally
\1
\mathrm{LHS}=\frac{1}{z}\left(\frac{\partial z}{\partial T}\right)_{V}\left[\rm{Li}_{1/2}\left(\eta z\right)+\phi\left(\kappa\Lambda\right)\rm{Li}_{-1/2}\left(\eta z\right)\right]-\frac{1}{T}\phi\left(\kappa\Lambda\right)\rm{Li}_{1/2}\left(\eta z\right)
\2
and
\1
\mathrm{RHS}=-\frac{3}{2T} \left[\rm{Li}_{3/2}\left(\eta z\right)+\phi\left(\kappa\Lambda\right)\rm{Li}_{1/2}\left(\eta z\right)\right]
\2
whence we obtain
\1
\frac{1}{z}\left(\frac{\partial z}{\partial T}\right)_{V,N}=-\frac{1}{2T}\frac{\left[3\rm{Li}_{3/2}\left(\eta z\right)+\phi\left(\kappa\Lambda\right)\rm{Li}_{1/2}\left(\eta z\right)\right]}{\left[\rm{Li}_{1/2}\left(\eta z\right)+\phi\left(\kappa\Lambda\right)\rm{Li}_{-1/2}\left(\eta z\right)\right]}.
\2

Substitution of the above expression in Eq. (\ref{spheat1}) and then use of Eq.(\ref{number}) finally yields expression for the specific heat as given in Eq.(\ref{specific heat}).

\setcounter{equation}{0}
\section{Fermionic chemical potential}
\label{appen-chempot}
In order to analyze analytically the existence of extrema in fermionic $\mu(T,\kappa)$, we use low-$T$ approximate expression
\1
\label{approx-mu}
\mu(T,\kappa)=\mu_0 \left[1 - \frac{\pi^2}{12}\left(\frac{k_{\rm B} T}{\mu_0}\right)^2 \frac{\mu_0 - \varepsilon_{\mathrm{so}}}{\mu_0 +  \varepsilon_{\mathrm{so}}} \right]
\2
available from Ref. \cite{hey:pra11}. The location of the Dirac point characterized by the energy $\varepsilon_{\mathrm{so}}$ increases monotonically whereas the zero-temperature chemical potential $\mu_0\equiv\mu(T=0,\kappa)$, i.e. the fermi energy $\varepsilon^0_{\mathrm{F}}$, decreases on increasing $\kappa$. The intersection of these two curves in the $\varepsilon-\kappa$ plane takes place at the critical value $\kappa_{\mathrm{c}}=2^{-2/3}k^0_\mathrm{F}$ which is the solution of $\mu_0= \varepsilon_{\mathrm{so}}$, i.e. $\mu(T=0,\kappa)=\hbar^2\kappa^2/(2m)$ with the expression for $\mu(T=0,\kappa)$ given in \cite{hey:pra11}. It is clear from Eq.(\ref{approx-mu})  that for $0\le\tilde\kappa<\tilde\kappa_{\mathrm{c}}$ (i.e. $\mu_0>\varepsilon_{\mathrm{so}}$), the curve bends down implying the maximum to be only at $T\g0$. On the other hand, the low-$T$ asymptote bends up for $\tilde\kappa>\tilde\kappa_{\mathrm{c}}$ implying that there is at least one maximum at finite $T$ because ultimately $\mu$ has to become negative at high $T$.

The detailed analysis of $\mu(T,\kappa)$ versus $T$ curves, generated based on computed values using the implicit relation for $\mu$ given in Eq.(\ref{number}),  for successively increasing $\kappa$ starting from zero reveals the following. Initially, the curve has normal behavior like an ideal Fermi gas without SOC but for $\tilde{\kappa}\approx0.5$, a shoulder begins to develop at $T\approx0.12 T_F^0$ which gets
more and more pronounced till $\tilde{\kappa}\approx0.5985$ but $\mu$ has its maximum at $T\g0$ for all these values of $\tilde{\kappa}$. On increasing $\kappa$, so far developing ``hanging shoulder" turns into an ``athletic shoulder"
showing a local minimum at $T_{\mathrm{min}}>0$ and a local maximum at $T_{\mathrm{max}}>T_{\mathrm{min}}$ apart from the global maximum at $T\g0$. We note that we get $\mu(T_{\mathrm{max}},\kappa)/\mu(0,\kappa)<1$ on the onset of max-min-max structure and this continues till the inequality gets reversed to $\mu(T_{\mathrm{max}},\kappa)/\mu(0,\kappa)>1$ for $\tilde{\kappa}\approx0.61$ without any change in order of max-min-max structure. The finite-$T$ minimum (maximum) wanders to lower (higher) $T$ as $\tilde{\kappa}$ is increased further. Finally, at $\tilde{\kappa}=\tilde\kappa_{\mathrm{c}}$, the finite-$T$ minimum gets merged with the zero-$T$ maximum leaving a local minimum at $T\g0$ and a global maximum at finite $T$. This structure continues for higher values of $\tilde{\kappa}$ but the location of the global maximum goes on shifting towards the vertical axis being depicted in Fig. \ref{fig-chempot}(a) as right portion (red dots) of the dotted curve.

What is the physical implication of $\pd{T}{\mu}>0$ for $\kappa>\kappa_{\mathrm{c}}$?
$\mu(T, \kappa)$ is the energy necessary to {\em add} one fermion to the system. For an ideal gas, this energy is the larger, the lower is $T$. It is the largest at $T\g0$, where it takes $\varepsilon_F^0$ to add an additional fermion to the system.
It takes so much energy even in the non--interacting gas, because there is the Pauli principle acting like an effective repulsion between particles. This statistics--induced repulsion is softened by SOC. To add a fermion at $T=0$ now will be the easier, the larger is $\kappa$. This statement also holds for any fixed $T$: $\mu(T, \kappa_1)>\mu(T, \kappa_2)$, if $\kappa_2>\kappa_1$.
On the contrary, for fixed $\kappa$, a corresponding statement does hold only  for $\kappa<\kappa_c$: $\mu(T_1, \kappa)>\mu(T_2, \kappa)$, if $T_2>T_1$ {\em and} $\kappa<\kappa_c$.

Obviously, the finite-$T$ maximum and its precursors (shoulder, min-max) result from a competition between the effects of  the statistical repulsion, which continuously increases with decreasing $T$, and the SOC, the effect of which is amplified by a factor $1/T$ and hence enormously at low $T$.
Also, for large $\kappa$ (fixed $T$), the distinction between bosons and fermions gets lost similar to what we find for large $T$ in ideal quantum gases without SOC.

\end{appendix}

\begin{thebibliography}{10}
\bibitem{dgj:rmp11}
J.~Dalibard, F.~Gerbier, G.~Juzeli\={u}nas, and P.~\"{O}hberg,
Rev. Mod. Phys. \textbf{83}, 1523 (2011).

\bibitem{xuy:pra12}
Z.~F. Xu and L.~You, Phys. Rev. A  \textbf{85}, 043605 (2012).

\bibitem{soc-rd}
E.~I. Rashba, Fiz. Tverd. Tela  \textbf{2}, 1224 (1960) [Sov. Phys. Solid State \textbf{2}, 1109 (1960)];
G.~Dresselhaus, Phys. Rev. \textbf{100}, 580 (1955).

\bibitem{zgz:pra13a}
Xiang-Fa Zhou, Guang-Can Guo, Wei Zhang, and Wei Yi,
Phys. Rev. A  \textbf{87}, 063606(R) (2013).

\bibitem{ljs:nat11}
Y.-J. Lin, K.~Jim\'{e}nez-Garc\'{\i}a, and I.~B. Spielman,
Nature \textbf{471}, 83 (2011).

\bibitem{soc-bose}
J.-Y. Zhang, S.-C. Ji, Z.~Chen, L.~Zhang, Z.-D. Du, B.~Yan, G.-S. Pan, B.~Zhao,
Y.-J. Deng, H.~Zhai, S.~Chen, and J.-W Pan,
Phys. Rev. Lett.  \textbf{109}, 115301  (2012);
Chunlei Qu, Chris Hamner, Ming Gong, Chuanwei Zhang, and Peter Engels,
arXiv:1301.0658v1, 2013.

\bibitem{soc-fermi}
P.~Wang, Z.-Q. Yu, Z.~Fu, J.~Miao, L.~Huang, S.~Chai, H.~Zhai, and J.~Zhang,
Phys. Rev. Lett. \textbf{109}, 095301  (2012);
L.~W. Cheuk, A.~T. Sommer, Z.~Hadzibabic, T.~Yefsah, W.~S. Bakr, and M.~W. Zwierlein.
\emph{ibid.}, 095302 (2012).

\bibitem{fhm:pra13}
Zhengkun Fu, Lianghui Huang, Zengming Meng, Pengjun Wang, Xia-Ji Liu, Han Pu,
Hui Hu, and Jing Zhang,
Phys. Rev. A  \textbf{87}, 053619 (2013).

\bibitem{ajg:prl12}
B.~M. Anderson, G.~Juzeli\={u}nas, V.~M. Galitski, and I.~B. Spielman,
Phys. Rev. Lett.  \textbf{108}, 235301  (2012).

\bibitem{lzw:prb12}
Yi~Li, Xiangfa Zhou, and Congjun Wu,
Phys. Rev. B  \textbf{85}, 125122  (2012).

\bibitem{weyl-fermion}
X.~Wan, A.~M. Turner, A.~Vishwanath, and S.~Y. Savrasov,
Phys. Rev. B  \textbf{83}, 205101  (2011);
A.~A. Burkov and Leon Balents, Phys. Rev. Lett.  \textbf{107}, 127205  (2011).

\bibitem{mag:05}
M.~Maggiore,
{\em A Modern Introduction to Quantum Field Theory}
(Oxford University Press, Oxford, 2005), p.55.

\bibitem{zlc:jpb13}
Xiangfa Zhou, Yi~Li, Zi~Cai, and Congjun Wu, J. Phys. B: At. Mol. Opt. Phys.  \textbf{46}, 134001 (2013).

\bibitem{reviews-soc}
Hui Zhai, Int. J. Mod. Phys. B  \textbf{26}, 1230001  (2012);
Victor Galitski and Ian~B. Spielman, Nature  \textbf{494}, 49 (2013).

\bibitem{half-vortex}
Subhasis Sinha, Rejish Nath, and Luis Santos,
Phys. Rev. Lett.  \textbf{107}, 270401 (2011);
Hui Hu, B.~Ramachandhran, Han Pu, and Xia-Ji Liu,
\emph{ibid.},  \textbf{108}, 010402 (2012).

\bibitem{striped}
Chunji Wang, Chao Gao, Chao-Ming Jian, and Hui Zhai,
Phys. Rev. Lett.  \textbf{105}, 160403 (2010);
Tin-Lun Ho and Shizhong Zhang,
\emph{ibid.}, \textbf{107}, 150403 (2011).

\bibitem{Fey:72}
R.~P. Feynman,
{\em Statistical Mechanics, A Set of Lectures}.
(Addison-Wesley Publishing Company, USA, 1972).

\bibitem{no-node}
Conjun Wu, Mod. Phys. Lett. B  \textbf{23}, 1 (2009);
Cong-Jun Wu, Ian Mondragon-Shem, and Xiang-Fa Zhou,
Chin. Phys. Lett.  \textbf{28}, 097102 (2011).

\bibitem{qu5:pra13}
Chunlei Qu, Chris Hamner, Ming Gong, Chuanwei Zhang, and Peter Engels,
Phys. Rev. A \textbf{88}, 021604(R) (2013).

\bibitem{gtz:prl11}
M.~Gong, S.~Tewari, and C.~Zhang,
Phys. Rev. Lett. \textbf{107}, 195303 (2011).

\bibitem{nontrivial}
Yi~Li and Congjun Wu, Phys. Rev. B  \textbf{85}, 205126 (2012);
Xiong-Jun Liu, K.~T. Law, and T.~K. Ng, arXiv:1304.0291v2, 2013.

\bibitem{vys:prb11}
J.~P. Vyasanakere and V.~B. Shenoy,
Phys. Rev. B  \textbf{83}, 094515 (2011).

\bibitem{hjl:prl11}
H.~Hu, L.~Jiang, X.-J. Liu, and H.~Pu,
Phys. Rev. Lett. \textbf{107}, 195304 (2011).

\bibitem{yuz:prl11}
Zeng-Qiang Yu and Hui Zhai,
Phys. Rev. Lett.  \textbf{107}, 195305 (2011).

\bibitem{majorana}
M.~Sato, Y.~Takahashi, and S.~Fujimoto,
Phys. Rev. Lett. \textbf{103}, 020401 (2009); Ming Gong, Gang Chen, Suotang Jia, and Chuanwei Zhang,
\emph{ibid.},  \textbf{109}, 105302 (2012);
X.-J. Liu and P.~D. Drummond,
Phys. Rev. A \textbf{86}, 035602 (2012);
Ran Wei1 and Erich~J. Mueller,
\emph{ibid.}, 063604 (2012);
Chunlei Qu, Zhen Zheng, Ming Gong, Yong Xu, Li~Mao, Xubo Zou, Guangcan Guo, and
Chuanwei Zhang, arXiv:1307.1207v1, 2013.

\bibitem{fflo}
Fan Wu, Guang-Can Guo, Wei Zhang, and Wei Yi,
Phys. Rev. Lett.  \textbf{110}, 110401 (2013);
Zhen Zheng, Ming Gong, Xubo Zou, Chuanwei Zhang, and Guangcan Guo,
Phys. Rev. A  \textbf{87}, 031602(R) (2013);
M.~Iskin and A.~L. Subasi, \emph{ibid.}, 063627 (2013); Xia-Ji Liu and Hui Hu, \emph{ibid.} \textbf{88}, 023622 (2013);
Zhen Zheng, Ming Gong, Yichao Zhang, Xubo Zou, Chuanwei Zhang, and Guangcan Guo,
arXiv:1212.6826v1, 2012;
Chunlei Qu, Zhen Zheng, Ming Gong, Yong Xu, Li~Mao, Xubo Zou, Guangcan Guo, and
Chuanwei Zhang, arXiv:1307.1207v1, 2013;
Wei Zhang and Wei Yi, arXiv:1307.2439v1, 2013;
Fan Wu, Guang-Can Guo, Wei Zhang, and Wei Yi, arXiv:1307.3117v1, 2013.

\bibitem{zhu6:pra13}
G.-B. Zhu, Q. Sun, Y.-Y. Zhang, K. S. Chan, W.-M. Liu, and A.-C. Ji,
Phys. Rev. A \textbf{88}, 023608 (2013).

\bibitem{reviews-cold}
Immanuel Bloch, Jean Dalibard, and Wilhelm Zwerger,
Rev. Mod. Phys.  \textbf{80}, 885 (2008);
S.~Giorgini, L.~P. Pitaevskii, and S.~Stringari, \emph{ibid.}, 1215 (2008);
M.~Inguscio, W.~Ketterle, and C.~Salomon, editors, {\em Proceedings of the
International School of Physics ``Enrico Fermi", Course CLXIV, Varenna, 20-30
June 2006} (IOS Press, Amsterdam, 2008);
Immanuel Bloch, Jean Dalibard, and Sylvain Nascimb$\grave{\mathrm{e}}$ne,
Nat. Phys.  \textbf{8}, 267 (2012).

\bibitem{lal:93}
L.~D. Landau and E.~M. Lifshitz,
{\em Statistical Mechanics, Part I}.
3rd Ed. (Pergamon Press, Oxford, UK, 1993).

\bibitem{soc:note4}
There is at least one exception to it. Spin-polarized $^1$H atoms are sometimes
referred to as ``spin-$\frac{1}{2}$ bosons", see, for example, A. J. Legget,
\emph{Quantum Liquids: Bose Condensation and Cooper pairing in
Condensed-Matter Systems}, (Oxford Univ. Press, New York, 2006), p.8.

\bibitem{pau:pr40}
W.~Pauli,
Phys. Rev.  \textbf{58}, 716 (1940).

\bibitem{cuz:pra13}
Xiaoling Cui and Qi~Zhou,
Phys. Rev. A  \textbf{87}, 031604(R) (2013).

\bibitem{hul:pra12}
H.~Hu and Xia-Ji Liu,
Phys. Rev. A  \textbf{85}, 013619 (2012).

\bibitem{ozb:prl12}
Tomoki Ozawa and Gordon Baym,
Phys. Rev. Lett. \textbf{109}, 025301 (2012).

\bibitem{hey:pra11}
Li~He and Zeng-Qiang Yu,
Phys. Rev. A  \textbf{84}, 025601 (2011).

\bibitem{vzs:prb11}
J.~P. Vyasanakere, S.~Zhang, and V.~B. Shenoy,
Phys. Rev. B \textbf{84}, 014512 (2011).

\bibitem{zhc:prl13}
Qi~Zhou and Xiaoling Cui,
Phys. Rev. Lett. \textbf{110}, 140407 (2013).

\bibitem{zyc:jpb13}
Wei Zheng, Zeng-Qiang Yu, Xiaoling Cui, and Hui Zhai,
J. Phys. B: At. Mol. Opt. Phys. \textbf{46}, 134007 (2013).

\bibitem{has:pra12}
Li~Han and C.~A.~R. S$\acute{a}$~de Melo,
Phys. Rev. A \textbf{85}, 011606(R) (2012).

\bibitem{yu:pra12}
Zhenhua Yu,
Phys. Rev. A \textbf{85}, 042711 (2012).

\bibitem{ghosh3:pra11}
Sudeep Kumar Ghosh, Jayantha P. Vyasanakere, and Vijay B. Shenoy,
Phys. Rev. A \textbf{84}, 053629 (2011).

\bibitem{anc:jpb13}
Brandon~M. Anderson and Charles~W. Clark,
J. Phys. B: At. Mol. Opt. Phys.  \textbf{46}, 134003 (2013).
Although the paper deals with the harmonically trapped gas with the SOC but it is due to isotropic trapping that the described
simplifying aspects would apply straightway in case of SO-coupled uniform gases too.

\bibitem{svg:cusc12}
Vijay~B. Shenoy, Jayantha~P. Vyasanakere, and S.~K. Ghosh,
Current Science  \textbf{103}, 525 (2012).

\bibitem{bagnato3:pra87}
Vanderlei Bagnato, David E. Pritchard and Daniel Kleppner,
Phys. Rev. A \textbf{35}, 4354(1987).

\bibitem{gradeshteyn2:book07}
I. S. gradeshteyn and I. M. Ryzhik,
{\em Table of integrals, Series and products}, 7th Ed.,
(Academic Press, New York 2007).

\bibitem{salasnich1:jmp00}
Luca Salasnich,
J. Math. Phys. \textbf{41}, 8016 (2000).

\bibitem{romero1:prl05}
See, for example, V\'{\i}ctor Romero-Roch\'{\i}n,
Phys. Rev. Lett. \textbf{94}, 130601 (2005); V\'{\i}ctor Romero-Roch\'{\i}n and Vanderlei S. Bagnato,
Brazilian J. Phys. \textbf{35}, 607 (2005).

\bibitem{pbm:book90}
A.~P. Prudnikov, Yu.~A. Brychkov, and O.~I. Marichev,
{\em Integrals and Series, Volume 3: More Special Functions}
(Gordon and Breach, New York, 1990).

\bibitem{lee:09}
M.~H. Lee,
Acta Physica Polonica B \textbf{40}, 1279 (2009).

\bibitem{pab:11}
R.~K. Patharia and Paul~D. Beale,
{\em Statistical Mechanics},
3rd ed. (Butterworth-Heinemann, Oxford, U.K., 2011).

\bibitem{gds:epjd03}
M.~Grether, M.~de~Llano, and M.~A. Sol\'{\i}s,
Eur. Phys. J. D \textbf{25}, 287 (2003).

\bibitem{sas:arxiv13}
P.~Salas and M.~A. Sol\'{\i}s,
arXiv:1307.2649v1, 2013.

\bibitem{pis:03}
L.~Pitaevskii and S.~Stringari,
{\em Bose--Einstein Condensation}
(Clarendon Press, Oxford, 2003), Chap. 7.

\bibitem{bps:pre11}
J.~Bosse, K.~N. Pathak, and G.~S. Singh,
Phys. Rev. E \textbf{84}, 042101 (2011).

\bibitem{brs:prl13}
A.~Blumkin, S.~Rinott, R.~Schley, A.~Berkovitz, I.~Shammass, and J.~Steinhauer,
Phys. Rev. Lett. \textbf{110}, 265301 (2013).


\end{thebibliography}

\end{document}